\documentclass[aps,floats]{revtex4}
\usepackage{amsmath,amssymb}
\usepackage{graphicx,epsfig}
\usepackage[greek,english]{babel}

\begin{document}
\bibliographystyle {plain}

\def\oppropto{\mathop{\propto}} 
\def\opsimeq{\mathop{\simeq}}
\def\opoverderline{\mathop{\overline}}
\def\operarrow{\mathop{\longrightarrow}}
\def\opsim{\mathop{\sim}}

\def\fig#1#2{\includegraphics[height=#1]{#2}}
\def\figx#1#2{\includegraphics[width=#1]{#2}}


\title{ Flow towards diagonalization for Many-Body-Localization models  :
\\ adaptation of the Toda matrix differential flow to random quantum spin chains } 


\author{ C\'ecile Monthus }
 \affiliation{Institut de Physique Th\'{e}orique, 
Universit\'e Paris Saclay, CNRS, CEA,
91191 Gif-sur-Yvette, France}

\begin{abstract}
The iterative methods to diagonalize matrices and many-body Hamiltonians can be reformulated as flows of Hamiltonians towards diagonalization driven by unitary transformations that preserve the spectrum. After a comparative overview of the various types of discrete flows (Jacobi, QR-algorithm) and differential flows (Toda, Wegner, White) that have been introduced in the past, we focus on the random XXZ chain with random fields in order to determine the best closed flow within a given subspace of running Hamiltonians. For the special case of the free-fermion random XX chain with random fields, the flow coincides with the Toda differential flow for tridiagonal matrices which is related to the classical integrable Toda chain and which can be seen as the continuous analog of the discrete QR-algorithm. For the random XXZ chain with random fields that displays a Many-Body-Localization transition, the present differential flow should be an interesting alternative to compare with the discrete flow that has been proposed recently to study the Many-Body-Localization properties in a model of interacting fermions (L. Rademaker and M. Ortuno, Phys. Rev. Lett. 116, 010404 (2016)).

\end{abstract}

\maketitle

\section{ Introduction }

In the field of Many-Body Localization  (see the recent reviews \cite{revue_huse,revue_altman} and references therein),
the focus is on the unitary dynamics of isolated random interacting quantum systems,
so that one needs to understand the properties of the whole set of excited eigenstates.
It is thus interesting to revisit the methods that have been proposed to diagonalize matrices and many-body Hamiltonians.

Whenever the eigenstates are not obvious, it is natural to devise iterative strategies.
For matrices, the idea to introduce an iterative method to converge towards diagonalization
goes back to the algorithm of Jacobi in 1846 \cite{jacobi} : the principle is to use iteratively elementary
two by two rotations in order to eliminate the corresponding off-diagonal terms.
This procedure has been adapted to many-body second-quantized Hamiltonians 
by White \cite{white} and has been applied recently to the problem
of Many-Body Localization for interacting fermions by Rademaker and Ortuno \cite{rademaker}.
Another possibility is to use continuous unitary transformations
as proposed independently by Wegner for condensed matter models \cite{wegner_first}
 (see the reviews \cite{wegner_reviews}, the book \cite{book_kehrein} and references therein), 
 by Glazek and Wilson for high-energy models \cite{wilson},
and by mathematicians for optimization problems \cite{math1,math2} under the name "double bracket flow".
In this continuous framework, various generators have been introduced : for instance the Wegner generator \cite{wegner_first,wegner_reviews,book_kehrein} eliminates more rapidly the off-diagonal terms corresponding to larger differences of the corresponding diagonal terms, while the White generator \cite{white} eliminates all off-diagonal terms with the same exponential rate. 

In all the schemes mentioned above, the goal is to obtain some systematic decrease of the off-diagonal part of the Hamiltonian.
However, there exists a completely different  strategy to converge towards diagonalization 
(see for instance the book \cite{book_krylov} and references therein ) : 
the "power method"  consists in the successive applications of the Hamiltonian $H$ onto some initial vector $\vert v_0>$
to converge towards the eigenvector associated to the biggest eigenvalue; the successive applications of the Hamiltonian can actually be kept as in the Lanczos algorithm; finally, instead of applying $H$ to a single vector,
the Hamiltonian can be applied to a basis of vectors $(\vert v_i >)$ in order to obtain the new vectors $H  \vert v_i >$ and 
 to produce a new basis after orthonormalization : this is the so-called QR-algorithm.
It turns out that the continuous formulation of this strategy displays
 very remarkable properties that have been much studied 
 under the name of "Toda flow" for tridiagonal matrices (see for instance \cite{henon,flaschka,moser,deift,symes,watkins,chu,elsner,tomei})
 as a consequence of its relation to the classical integrable Toda lattice \cite{toda}.
Note that exactly the same generator of continuous unitary transformations
has been re-discovered independently by Mielke \cite{mielke_band}
via the requirement to obtain a closed flow for band matrices, i.e. to maintain the 'sparsity'
of the initial matrix. 

For Many-Body Hamiltonians, the idea
 to avoid that the diagonalization flow invades the whole space of possible running Hamiltonians
is of course even more essential.
 The goal of the present paper is thus to try to adapt the strategy of
Toda differential flow of tridiagonal matrices to random quantum spin chains.
However since the literature
 on the various methods described above is scattered over various communities,
with various ideas re-invented several times independently, it seems useful to give first some comparative overview of the different frameworks. 
 
The paper is organized as follows.
In Section \ref{sec_flow}, the notion of flow towards diagonalization via unitary transformations
is presented with its invariants, both for matrices and for Many-Body quantum spin chains.
For matrices, discrete flows (Jacobi, QR) are described in Section \ref{sec_discretematrix},
while the continuous flows  (Wegner, White, Toda) are recalled in 
Section \ref{sec_continuousmatrix}.
For random quantum spin chains, the discrete flow translated from the corresponding 
interacting fermions formulation (White, Rademaker-Ortuno) is presented
in Section \ref{sec_discreteMB}, while the continuous framework is given
in Section \ref{sec_continuousMB}.
In section \ref{sec_todaspin}, we adapt the idea of the Toda flow for the XXZ chain.
Our conclusions are summarized in section \ref{sec_conclusion}.

\section{ Notion of flow towards diagonalization }

\label{sec_flow}

The goal is to diagonalize the Hamiltonian $H$,
i.e. to find the eigenvalues $E_{i}$ and the corresponding eigenstates 
$\vert \psi_i>$ 
\begin{eqnarray}
 H = \sum_i E_{i} \vert \psi_i> < \psi_i \vert
\label{Hspectrum}
\end{eqnarray}
via an iterative procedure based on unitary transformations.

\subsection{ Family of unitary transformations }

Let $l$ be an index that can be either discrete $l=0,1,2,..$
or continuous $l \in [0,+\infty[ $.
One wishes to construct a series of unitary transformation $U(l)$
\begin{eqnarray}
U(l) U^{\dagger}(l) = U^{\dagger}(l)U(l)=1
\label{unitary}
\end{eqnarray}
so that the running Hamiltonian 
\begin{eqnarray}
H(l)=U(l) H U^{\dagger}(l) 
\label{Hl}
\end{eqnarray}
starting at $H(l=0)=H$ converges towards diagonalization as $l \to +\infty$.
Note that this goal is very ambitious for complex models, since all information is kept along the flow, in contrast to renormalization methods that try to eliminate iteratively the irrelevant information.

\subsection{ Invariants of the flow }

Since the flow built from unitary transformations
conserves the eigenvalues $E_i$,
it is interesting to construct the invariants 
corresponding to the traces of the integer powers of the Hamiltonian
\begin{eqnarray}
I_p \equiv  \sum_i E_i^p = {\rm Tr}  (H^p(l)) 
\label{In}
\end{eqnarray}
The conservation of the invariant for $p=1$ leads to
the sum rule of the diagonal elements $H_{n,n}(l) $ 
in terms of the energies $E_i$
\begin{eqnarray}
I_1 \equiv  \sum_i E_i = Tr (H(l)) = \sum_n H_{n,n}(l)
\label{I1}
\end{eqnarray}

The conservation of the invariant for $p=2$ leads to
the sum rule for the square of the modulus of the matrix elements
\begin{eqnarray}
I_2  \equiv  \sum_i E_i^2 = Tr (H^2(l)) = \sum_{n} \sum_{m} H_{n,m}(l) H_{m,n}(l) 
 =  \sum_{n} \sum_{m} \vert H_{n,m}(l) \vert^2
\label{I2}
\end{eqnarray}
This means that its diagonal and off-diagonal contributions 
\begin{eqnarray}
I_2 && =I_2^{diag}(l) +I_2^{off}(l) 
\nonumber \\
I_2^{diag}(l)&& \equiv \sum_{n}   H_{n,n}^2(l)
\nonumber \\
I_2^{off}(l) &&\equiv  \sum_{n} \sum_{m\ne n }   \vert  H_{n,m}(l) \vert^2
\label{I2diagoff}
\end{eqnarray}
have opposite variations
\begin{eqnarray}
\frac{dI_2^{diag}(l)}{dl} = - \frac{dI_2^{off}(l)}{dl}   
\label{I2diagoffopp}
\end{eqnarray}
The goal of full diagonalization  at $l=+\infty$ corresponds to
\begin{eqnarray}
I_2^{off}(l=+\infty) && =0
\nonumber \\
I_2^{diag}(l=+\infty) && = I_2 =  \sum_i E_i^2
\label{I2diagoffinal}
\end{eqnarray}

This second invariant allows to distinguish between two types of strategies :

(i) either one imposes that
the off-diagonal contribution $I_2^{off}(l) $ is always decaying
from its initial value $I_2^{off}(l=0) $ towards its vanishing final value
$I_2^{off}(l=+\infty)=0$
\begin{eqnarray}
 \frac{dI_2^{off}(l)}{dl}   < 0
\label{I2diagoffflow}
\end{eqnarray}

(ii) or the only condition imposed on the dynamics of 
off-diagonal contribution $I_2^{off}(l) $ is its final vanishing final value
$I_2^{off}(l=+\infty)=0$. To obtain this long-term objective, one is ready
to accept a temporary increase of $I_2^{off}(l) $ along the flow.

\subsection{ Application to Many-Body Hamiltonians }

For many-body Hamiltonians, one can  
consider that the Hamiltonian is represented by a matrix in a given basis of
the Hilbert space and apply the methods developed for matrices.
However it seems much more appropriate to define the diagonalization flow 
in terms of the coupling constants in front of second-quantized operators.

For instance for a chain of $N$ quantum spins described by 
 the hermitian Pauli matrices at each site
\begin{eqnarray}
\sigma^{(0)}  = Id =
 \begin{pmatrix}
1 & 0 \\
0 & 1 
\end{pmatrix}
\ \ \ \ \ 
\sigma^{x} 
= \begin{pmatrix}
0 & 1 \\
1 & 0 
\end{pmatrix}
\ \ \ \ \ 
\sigma^{y}  =
 \begin{pmatrix}
0 & -i \\
i & 0 
\end{pmatrix}
\ \ \ \ \ 
\sigma^{z} && =
 \begin{pmatrix}
1 & 0 \\
0 & -1 
\end{pmatrix}
\label{pauli}
\end{eqnarray}
 the most general running Hamiltonian for $N$ spins can be expanded 
in the basis of Pauli matrices as
\begin{eqnarray}
H(l) =  \sum_{a_1=0,x,y,z} ...  \sum_{a_N=0,x,y,z} 
H_{a_1...a_N }(l)   \sigma_1^{(a_1)}  \sigma_2^{(a_2)}  ...  \sigma_N^{(a_N)} 
\label{hl}
\end{eqnarray}
with the $4^N$ real coefficients
\begin{eqnarray}
H_{a_1...a_N }(l)  = \frac{1}{2^N} Tr (H(l)  \sigma_1^{(a_1)}  \sigma_2^{(a_2)}  ...  \sigma_N^{(a_N)}   ) 
\label{hail}
\end{eqnarray}

The diagonal part 
\begin{eqnarray}
H_{diag}(l) =  \sum_{a_1=0,z} ...  \sum_{a_N=0,z}
H_{a_1...a_N }(l)   \sigma_1^{(a_1)}  \sigma_2^{(a_2)}  ...  \sigma_N^{(a_N)} 
\label{hldiag}
\end{eqnarray}
involves the $2^N$ real coefficients $H_{a_1...a_N }(l)  $,
where each $a_i$ takes only the two values $a_i=0,z$.
The goal is to flow towards this diagonal form, and thus to eliminate
the off-diagonal part
\begin{eqnarray}
H_{off}(l)= H(l) - H_{diag}(l) 
\label{hloff}
\end{eqnarray}
that contains some Pauli matrices $ \sigma_i^{a_i} $ with the values $a_i=x,y$.

Here it is convenient to normalize the invariants as
\begin{eqnarray}
I_q=  \frac{1}{2^N} Tr (H^q(l)) 
\label{inviq}
\end{eqnarray}
The first invariant $q=1$ given by the coefficient of Eq. \ref{hail}
where all indices vanish $a_i=0$
\begin{eqnarray}
I_1=\frac{1}{2^N} Tr(H(l)) =  H_{0,0,0,..0 }(l) 
\label{hli1}
\end{eqnarray}
simply represents the middle of the spectrum and can be chosen to vanish.

The second invariant corresponds to the sum of the squares of all coefficients
of Eq. \ref{hl}
\begin{eqnarray}
I_2= \frac{1}{2^N} Tr(H^2(l)) = \sum_{a_1=0,x,y,z} ...  \sum_{a_N=0,x,y,z} H^2_{a_1...a_N }(l)
\label{hli2}
\end{eqnarray}
with the diagonal and off-diagonal contributions
\begin{eqnarray}
I_2^{diag}(l) && =\frac{1}{2^N} Tr( H^2_{diag}(l)  ) = \sum_{a_1=0,z} ...  \sum_{a_N=0,z} H^2_{a_1...a_N }(l)
\nonumber \\
I_2^{off}(l) && =I_2 - I_2^{diag}(l)
\label{hli2diag}
\end{eqnarray}

\subsection{ Example of the XXZ chain with random fields  }

The quantum spin chains with local interactions
are of course extremely
 sparse with respect to the space of couplings of dimension
$4^N$ discussed above.
For instance, in the XXZ chain with random couplings and random fields,
where the diagonal and off-diagonal parts read
(the XX coupling is defined with respect
to the ladder operators $\sigma^{\pm}=(\sigma^x \pm \sigma^y)/{2} $
for later convenience)
\begin{eqnarray}
H_{diag}^{XXZ}(l=0) && = \sum_{i=1}^N ( h_i \sigma_i^z + J^{zz}_i  \sigma_i^z \sigma_{i+1}^z )
\nonumber \\
H_{off}^{XXZ}(l=0) &&
=  \sum_{i=1}^N \frac{J_i}{2} ( \sigma_i^x \sigma_{i+1}^x+\sigma_i^y \sigma_{i+1}^y )
= \sum_{i=1}^N J_i ( \sigma_i^+ \sigma_{i+1}^-+\sigma_i^- \sigma_{i+1}^+ )
\label{XXZh}
\end{eqnarray}
the only non-vanishing coefficients of Eq. \ref{hail} are
the $N$ terms with only one non-vanishing index $a_i=z$,
and the $(3N)$ terms with only two consecutive coinciding indices $a_i=a_{i+1} \ne 0$
\begin{eqnarray}
H_{a_1...a_N }(l=0) && = h_i \delta_{a_i,z} \prod_{j \ne i} \delta_{a_j,0}
+ J_i^{zz} \delta_{a_i,z} \delta_{a_{i+1},z} \prod_{j \ne (i,i+1)} \delta_{a_j,0}
+ \frac{J_i}{2} ( \delta_{a_i,x} \delta_{a_{i+1},x} +
\delta_{a_i,y} \delta_{a_{i+1},y}) \prod_{j \ne (i,i+1)} \delta_{a_j,0}
\label{hailheisen}
\end{eqnarray}
The first invariant of Eq. \ref{hli1}
vanishes while the second invariant of Eq. \ref{hli2diag} 
\begin{eqnarray}
I_2 && = I_2^{diag}(l=0)+I_2^{off}(l=0)
\nonumber \\
I_2^{diag}(l=0) && = \sum_i (h_i^2 + (J_i^{zz})^2   ) 
\nonumber \\
I_2^{off}(l=0) && = \frac{1}{2} \sum_i J_i^2
\label{hli2diagzero}
\end{eqnarray}
grows linearly in $N$ and fixes the variance of
the Gaussian form of the density of states
in the middle of the spectrum as in other local spin models \cite{atas,keating}.

Via the standard Jordan-Wigner transformation onto anticommuting fermionic operators
\begin{eqnarray}
c_i^{\dagger} && \equiv e^{i \pi \sum_{j=1}^{i-1} \sigma_j^+ \sigma_j^- }  \sigma_i^+
\nonumber \\
c_i && \equiv e^{- i \pi \sum_{j=1}^{i-1} \sigma_j^+ \sigma_j^- }  \sigma_i^-
\nonumber \\
c_i^{\dagger}c_i && = \sigma_i^+\sigma_i^-
\label{jordanwigner}
\end{eqnarray}
the XXZ Hamiltonian of Eq. \ref{XXZh} becomes
\begin{eqnarray}
H_{diag}^{XXZ}(l=0) && = \sum_{i=1}^N  h_i ( 2c_i^{\dagger}c_i -1) +
 \sum_{i=1}^NJ^{zz}_i  ( 2c_i^{\dagger}c_i -1) ( 2c_{i+1}^{\dagger}c_{i+1} -1)
+ \sum_{i=1}^N J_i ( c_i^{\dagger}c_{i+1} + c_{i+1}^{\dagger}c_{i} )
\label{XXZfermions}
\end{eqnarray}

The $XX$ chain corresponding to $J^{zz}=0$ is a free-fermion quadratic model
that can be diagonalized into $H^{free}= \sum \epsilon_k f_k^{\dagger} f_k$
via a unitary transformation of the fermions operators from $c_i$ to $f_k$, i.e. in the spin language, the diagonalization form obtained by the diagonalization flow at $l=+\infty$ should contain only $N$ coefficients $h_i(l=+\infty) $ 
\begin{eqnarray}
H^{XX}(l=+\infty) && = \sum_{i=1}^N  h_i(l=+\infty) \sigma_i^z
\label{XXflotinfinity}
\end{eqnarray}
As a consequence, our first goal in this paper
is to define a flow 
that remain sparse for the free-fermion XX chain between
 the sparse initial state of Eq. \ref{XXZh} and the sparse final state of Eq. \ref{XXflotinfinity}.

For the XXZ chain with $J^{zz} \ne 0 $, there exists a non-trivial 
density-density interaction in the fermion language of Eq. \ref{XXZfermions},
so that one does not expect any simplification with respect to the most general
diagonal form of Eq. \ref{hldiag}
at $l=+\infty$
\begin{eqnarray}
H^{XXZ}(l=+\infty) =  \sum_{a_1=0,z} ...  \sum_{a_N=0,z}
H_{a_1...a_N }(l=+\infty)   \sigma_1^{(a_1)}  \sigma_2^{(a_2)}  ...  \sigma_N^{(a_N)} 
\label{hldiagxxzinfty}
\end{eqnarray}
where the $2^N$ real coefficients are necessary to reproduce the $2^N$ 
eigenvalues. 
The form of Eq. \ref{hldiagxxzinfty} has been much discussed
in the context of Many-Body Localization 
 (see the recent reviews \cite{revue_huse,revue_altman} and references therein).
In particular, in the Fully-Many-Body-Localized phase, 
the pseudo-spin operators at $l=+\infty$
that commute with each other and with the Hamiltonian
are related to the initial spins by a {\it quasi-local } unitary transformation
and the couplings $H_{a_1...a_N }(l=+\infty)$ decay exponentially with the distance with a sufficient rate so that there exists an extensive number of emergent {\it localized } conserved operators (see \cite{emergent_swingle,emergent_serbyn,emergent_huse,emergent_ent,imbrie,serbyn_quench,emergent_vidal,emergent_ros,rademaker} and the reviews \cite{revue_huse,revue_altman} for more details).
In particular, these emergent conserved operators have a simple interpretation
 within the RSRG-X procedure that constructs 
the whole set of excited eigenstates
 \cite{rsrgx,rsrgx_moore,vasseur_rsrgx,yang_rsrgx,rsrgx_bifurcation,emergent_c}
via the extension of the Strong Disorder Real-Space RG approach
developed by Ma-Dasgupta-Hu \cite{ma_dasgupta}
 and Daniel Fisher \cite{fisher_AF,fisher}
to construct the ground states
(see \cite{fisherreview,strong_review,refael_review} for reviews).
The Heisenberg chain in random fields $h_i$
corresponding to the uniform couplings $J^{zz}_i=\frac{J_i}{2}=J$
is the model displaying a Many-Body-Localization transition 
that has been studied on the biggest sizes (see \cite{alet,alet_dyn,luitz}
and references therein).

For the XXZ chain, the final state of Eq. \ref{hldiagxxzinfty}
thus contains $2^N$ coefficients, in contrast to the initial state containing only
$4N$ coefficients (Eq \ref{hailheisen}). So here our goal 
will be to obtain a diagonalization flow $H_{XXZ}(l)$
that remain sparse with respect to the ladder operators.

\section{ Discrete diagonalization flows for matrices   }

\label{sec_discretematrix}

 \subsection{ Generator $\eta(l)$ associated to the elementary unitary transformations
 $u (l)$  }

When $l$ is a discrete index, the unitary transformations $U(l)$ are constructed
via some iteration
\begin{eqnarray}
U(l) \equiv u(l) U(l-1) = u(l) u(l-1) u(l-2) ... u(1)
\label{unitaryrec}
\end{eqnarray}
in terms of elementary unitary transformations $u(l)$
that governs the flow of the running Hamiltonian of Eq. \ref{Hl}
\begin{eqnarray}
H(l)=U(l) H U^{\dagger}(l) = u(l) H(l-1) u^{\dagger}(l)
\label{Hlflow}
\end{eqnarray}

It is convenient to introduce the anti-hermitian generator $\eta(l)$ 
\begin{eqnarray}
\eta^{\dagger} (l) = - \eta(l)
\label{etaanti}
\end{eqnarray}
associated
to the elementary unitary transformations $u(l)$
\begin{eqnarray}
u(l) \equiv e^{\eta(l)}
\label{etau}
\end{eqnarray}

\subsection{ Jacobi's algorithm for matrices  }

\label{sec_jacobi}

In Jacobi's algorithm \cite{jacobi}, 
one identifies the off-diagonal element with the biggest modulus
of the matrix $H(l-1)$
\begin{eqnarray}
\vert H_{mn}(l-1) \vert = {\rm max}_{i<j} \vert H_{ij}(l-1) \vert
\label{offmax}
\end{eqnarray}
The antihermitian generator of Eq. \ref{etauanti} is chosen as
\begin{eqnarray}
 \eta(l) =\theta_l \left(  e^{i \phi_l}  \vert n > <m  \vert  - e^{-i \phi_l} \vert m > <n  \vert \right)
\label{etauanti}
\end{eqnarray}
in order to produce the complex unitary rotation
\begin{eqnarray}
u(l) && =  e^{\eta(l)}
= 1 + (\cos \theta_l -1) (\vert m > <m  \vert + \vert n > <n  \vert)
+ \sin \theta_l ( e^{i \phi_l}\vert n > <m  \vert  -e^{-i \phi_l}\vert m > <n  \vert  ) 
\label{urot}
\end{eqnarray}
 acting in the two-dimensional subspace $(m,n)$ as
\begin{eqnarray}
 u(l) \vert m > && = \cos \theta_l  \vert m>
 +  e^{i \phi_l} \sin \theta_l  \vert n>
\nonumber \\
u(l) \vert n > && = - e^{-i \phi_l} \sin \theta_l  \vert m> 
+ \cos \theta_l \vert n> 
\label{urotation}
\end{eqnarray}

The two angles $\theta_l$ and $\phi_l$ are chosen to make the off-diagonal element
of $H(l)$ between $m$ and $n$ vanish 
\begin{eqnarray}
0 &&  =H_{mn}(l) = < m \vert u(l) H(l-1)  u^{\dagger} (l) \vert n>
\label{etaulin0}
\end{eqnarray}
 yielding
\begin{eqnarray}
0 &&  =\cos^2 \theta_l e^{i \phi_l} H^*_{nm} (l-1)- \sin^2 \theta_l e^{- i \phi_l}H_{nm} (l-1)
+ \cos \theta_l \sin \theta_l \left( H_{mm} (l-1)-  H_{nn} (l-1)\right)
\label{etaulin00}
\end{eqnarray}
The phase $e^{i \phi_l}$ is thus the phase
 of the off-diagonal element $H_{nm} (l-1) $
\begin{eqnarray}
e^{i \phi_l} && = \frac{ H_{nm}(l-1) }{\vert  H_{nm}(l-1)  \vert}
\label{phil}
\end{eqnarray}
while the remaining real equation for the angle $\theta_l$ 
\begin{eqnarray}
0 &&  = \cos (2 \theta_l ) \vert H_{nm} (l-1) \vert
+ \sin (2 \theta_l) \left( \frac{H_{mm} (l-1)-  H_{nn} (l-1)}{2} \right)
\label{etaulin000}
\end{eqnarray}
is determined by the ratio between the off-diagonal element modulus
and the difference between the diagonal elements
\begin{eqnarray}
\tan (2 \theta_l) && =  \frac{ 2 \vert H_{nm} (l-1) \vert }{
 H_{nn} (l-1)-  H_{mm} (l-1) }
\label{etaulin}
\end{eqnarray}

The new diagonal elements correspond to the standard formula for the 
diagonalization of $2 \times 2$ matrices
\begin{eqnarray}
H_{mm}(l) 
&& = \frac{ H_{mm}(l-1) + H_{nn}(l-1)}{2}
+ \frac{ H_{mm}(l-1) - H_{nn}(l-1)}{2} \sqrt{1+ \left(  \frac{ 2 \vert H_{nm} (l-1) \vert }{
 H_{nn} (l-1)-  H_{mm} (l-1) } \right)^2}  
\nonumber \\
H_{nn}(l) && = \frac{ H_{nn}(l-1) + H_{mm}(l-1)}{2}
+ \frac{ H_{nn}(l-1) - H_{mm}(l-1)}{2} \sqrt{1+ \left(  \frac{ 2 \vert H_{nm} (l-1) \vert }{
 H_{nn} (l-1)-  H_{mm} (l-1) } \right)^2}  
\label{jacobiflowdiag}
\end{eqnarray}
The sum is conserved as it should (Eq. \ref{I1})
\begin{eqnarray}
I_1(l)-I_1(l-1) && = H_{mm}(l)+ H_{nn}(l)- H_{mm}(l-1)- H_{nn}(l-1) 
= 0
\label{I1diagjac}
\end{eqnarray}
while the diagonal contribution to the second invariant of Eq. \ref{I2diagoff}
evolves according to
\begin{eqnarray}
&& I_2^{diag}(l)-I_2^{diag}(l-1)  = H_{mm}^2(l)+ H_{nn}^2(l)- H_{mm}^2(l-1)- H_{nn}^2(l-1)  = 2  \vert H_{nm}(l-1) \vert^2
\label{I2diagjac}
\end{eqnarray}
So the maximum rule of Eq. \ref{offmax} corresponds to the maximal growth
of the diagonal contribution, and thus to the maximal decay of the off-diagonal
contribution (Eq. \ref{I2diagoff}) among the choice of
elementary $2 \times 2$ rotations.
Note however that the choice of the maximal off-diagonal element in Eq. \ref{offmax}
is not mandatory : one can choose a different order among the off-diagonal elements
if it is more convenient for practical reasons, 
and the convergence towards diagonalization will be still ensured by
Eq. \ref{I2diagjac}.

\subsection{ QR algorithm for matrices }

\label{sec_qr}

In the Jacobi algorithm described above, 
the goal is to obtain the systematic decrease of the off-diagonal part of the Hamiltonian.
However there exist completely different strategies to converge towards diagonalization 
(see for instance the book \cite{book_krylov} and references therein).
For instance, the convergence towards the eigenvector associated with the biggest eigenvalue
can be achieved by the successive applications $H^n \vert v_0>$ of the Hamiltonian $H$
 onto some initial vector $\vert v_0>$, this is the so-called "power method".
To obtain more eigenvectors, the successive applications
 of the Hamiltonian can be kept to build the Krylov subspace spanned by the successive iterations 
$(  \vert v_0>, H \vert v_0>, ..., H^n \vert v_0> )$ and one orthogonal basis can be constructed via the Lanczos algorithm.
 When the goal is the full diagonalization, these ideas can be used slightly differently
to obtain the so-called QR-algorithm : 
 instead of applying $H$ to a single vector,
the Hamiltonian can be applied to the current basis $\vert i >$ of vectors 
to obtain the new vectors $H  \vert i >$ 
in order to produce a new basis after orthonormalization : this amounts to
the so-called QR-decomposition of the current Hamiltonian $H(l)$ into 
\begin{eqnarray}
H(l) = Q(l) R(l)
\label{qrdecomposition}
\end{eqnarray}
where $Q(l)$ is the orthogonal matrix ($Q(l) Q^t(l) =Id$) describing the change of bases
and where $R(l)$ is the upper triangular matrix produced by the orthonormalization
process of the vectors $(H  \vert i > )$ (for instance via the Gram-Schmidt procedure
or via the Householder reflections method).
Then the writing of the Hamiltonian in the new basis gives the iteration
\begin{eqnarray}
H(l+1) =Q^t (l) H(l)  Q(l) = Q^t (l) ( Q(l) R(l) ) Q(l) = R(l) Q(l)
\label{qralgo}
\end{eqnarray}
This defines the QR-algorithm that converges towards diagonalization.

In summary, the QR-algorithm is an important example of a strategy
not driven by the systematic decay of the off-diagonal elements via $I_2^{off}$,
but based instead on the long-term expectation that the successive applications
of the Hamiltonian will converge towards diagonalization asymptotically.

\section{ Continuous diagonalization flows for matrices   }

\label{sec_continuousmatrix}

\subsection{ Generator $\eta(l)$ of the infinitesimal unitary transformation}

When the family of unitary transformation $U(l)$
is parametrized by a continuous parameter $l \in [0,+\infty[ $,
 the elementary unitary transformation to go from $l$ to $(l+dl)$
has some infinitesimal amplitude $(dl)$ in front of the anti-Hermitian
generator $\eta(l)$
\begin{eqnarray}
U(l+dl) && = e^{dl \eta(l)} U(l) = U(l) + dl \eta(l) U(l)
\label{etacontinuous}
\end{eqnarray}

Then the running Hamiltonian $H(l)$ evolves according to
\begin{eqnarray}
H(l+dl) =  e^{dl \eta(l)} H(l)  e^{-dl \eta(l)} = H(l) + dl (\eta(l) H(l) - H(l) \eta(l)  )
\label{Adl}
\end{eqnarray}
i.e. it satisfies the differential equation involving the commutator with the generator
\begin{eqnarray}
\frac{dH(l)}{dl}= \left[ \eta(l), H(l) \right]
\label{deriHl}
\end{eqnarray}

In terms of the matrix elements of the anti-hermitian generator 
\begin{eqnarray}
\eta_{nk}(l) && = - \eta^*_{k,n}(l)
\label{etankanti}
\end{eqnarray}
the off-diagonal matrix elements $n \ne q$ evolve according to
\begin{eqnarray}
\frac{dH_{nq}(l)}{dl} && = - (H_{nn}(l)-H_{qq}(l) ) \eta_{nq}(l)
+ \sum_{k \ne (n,q)} \left( \eta_{nk}(l) H_{kq}(l) - H_{nk}(l) \eta_{kq}(l) \right)
\label{deriHlbasispq}
\end{eqnarray}
while the evolution of the diagonal elements reads
\begin{eqnarray}
\frac{dH_{nn}(l)}{dl} && =\sum_{k \ne n} \left( \eta_{nk}(l) H_{kn}(l)
 +  \eta_{nk}^*(l) H_{kn}^*(l)  \right)
\label{deriHlbasisnn}
\end{eqnarray}

The flow of the off-diagonal contribution $I_2^{off}(l) $
and of the diagonal contribution $I_2^{diag}(l)$ of Eq. \ref{I2diagoff} becomes
\begin{eqnarray}
 - \frac{dI_2^{off}(l)}{dl}  = \frac{dI_2^{diag}(l)}{dl} 
&& = 2 \sum_{n}   H_{nn}(l)\frac{d H_{nn}(l)}{dl} 
\nonumber \\
&& =  \sum_{n \ne k}  ( H_{nn}(l)- H_{kk}(l) )  \left( \eta_{nk}(l) H_{kn}(l) +\eta_{nk}^*(l) H_{kn}^*(l)  \right)
\label{I2diagflowconti}
\end{eqnarray}

\subsection{ Flows based on the systematic decay of $I_2^{off}(l)$ }

Generators of the form
\begin{eqnarray}
\eta_{nk}= H_{nk} f(H_{nn}-H_{kk})
\label{etaf}
\end{eqnarray}
where the function $f(x)$ is antisymmetric $f(x)=-f(-x)$ to insure  
the antihermitian condition of Eq. \ref{etankanti} and satisfies $x f(x) \geq 0$
will produce flows with a systematic decay of $I_2^{off}(l) $  (Eq. \ref{I2diagflowconti})
\begin{eqnarray}
0< - \frac{dI_2^{off}(l)}{dl}  = \frac{dI_2^{diag}(l)}{dl} 
&& = 2  \sum_{n \ne k} \vert  H_{nk}(l) \vert^2  ( H_{nn}(l)- H_{kk}(l) ) f( H_{nn}(l)- H_{kk}(l) )
\label{I2diagflowcontif}
\end{eqnarray}

\subsubsection{ Wegner's choice $f^{Wegner}(x)  =x$   }

Wegner \cite{wegner_first,wegner_reviews,book_kehrein}
has proposed to choose as generator
\begin{eqnarray}
\eta^{Wegner}_{nk}(l) && = H_{nk}(l) (H_{nn}(l) - H_{kk}(l) )
\label{etawegnernk}
\end{eqnarray}
As the level of operators, these matrix elements correspond
to the commutator between
the diagonal part and the off-diagonal part of the Hamiltonian
\begin{eqnarray}
 \eta^{Wegner}(l) = \left[H_{diag}(l) , H(l) \right] = \left[H_{diag}(l) , H_{off}(l) \right]
\label{etawegner}
\end{eqnarray}

Then $I_2^{off}(l)$ is a decreasing function (Eq. \ref{I2diagflowconti})
\begin{eqnarray}
0< - \frac{dI_2^{off}(l)}{dl}  = \frac{dI_2^{diag}(l)}{dl} 
&& =  \sum_{n \ne k}  ( H_{nn}(l)- H_{kk}(l) )^2  \vert H_{nk} \vert^2
\label{I2diagflowcontiwegner}
\end{eqnarray}
The off-diagonal matrix elements $n \ne q$ (Eq \ref{deriHlbasispq})
\begin{eqnarray}
\frac{dH_{nq}(l)}{dl} && = - (H_{nn}(l)-H_{qq}(l) )^2 H_{nq}(l)
+ \sum_{k \ne (n,q)} H_{nk}(l)  H_{kq}(l)
 \big(  H_{nn}(l)+  H_{qq}(l) - 2 H_{kk}(l) \big)
\label{deriHlbasispqwegner}
\end{eqnarray}
are suppressed more rapidly when they are
associated to large differences of diagonal elements $( H_{nn}(l)- H_{qq}(l) )^2 $.
The equations of motion for the diagonal elements  (Eq. \ref{deriHlbasisnn}) 
\begin{eqnarray}
\frac{dH_{nn}(l)}{dl} && =2 \sum_{k \ne n} \vert H_{nk}(l) \vert^2 (H_{nn}(l) - H_{kk}(l) )
\label{deriHlbasisnnwegner}
\end{eqnarray}
and for the off-diagonal elements of Eq. \ref{deriHlbasispqwegner}
are polynomial of degree 3 in the matrix elements.

\subsubsection{ White's choice $f^{White}(x) = \frac{1}{x}$    }

White has proposed the generator (Eq. 18 of Reference \cite{white})
\begin{eqnarray}
\eta^{White}_{nk}(l) && = \frac{H_{nk}(l) }{H_{nn}(l) - H_{kk}(l) }
\label{etawhitenk}
\end{eqnarray}
Then the dynamics of $I_2^{off}(l) $ (Eq. \ref{I2diagflowconti})
\begin{eqnarray}
0< - \frac{dI_2^{off}(l)}{dl} 
= 2  \sum_{n \ne k} \vert H_{kn}(l) \vert^2 = 2 I_2^{off}(l)
\label{I2diagflowwhite}
\end{eqnarray}
has the nice property to display the explicit exponential decay
\begin{eqnarray}
 I_2^{off}(l) = I_2^{off}(l=0) e^{- 2 l}
\label{I2diagflowwhiteexp}
\end{eqnarray}
(The particular value $2$ of the exponential decay $e^{-2 l}$
 has no particular meaning and can be changed by a 
redefinition of the flow parameter $l$).

The choice of Eq. \ref{etawhitenk} can be seen as the infinitesimal
counterpart of the Jacobi choice of Eq. \ref{etaulin} 
and of the convergence rate towards
diagonalization of Eq. \ref{I2diagjac} : the difference is that instead of doing
a single finite elementary rotation to eliminate the biggest off-diagonal element,
one uses the commutativity of all 
infinitesimal rotations to make all the off-diagonal
elements decay with the same rate.

The corresponding equations of motion for the 
off-diagonal matrix elements $n \ne q$ (Eq \ref{deriHlbasispq})
\begin{eqnarray}
\frac{dH_{nq}(l)}{dl} &&
  = - H_{nq}(l)
+ \sum_{k \ne (n,q)}H_{nk}(l)H_{kq}(l) \left( \frac{1 }{H_{nn}(l) - H_{kk}(l) }
 -  \frac{ 1 }{H_{kk}(l) - H_{qq}(l) }\right)
\label{deriHlbasispqwhite}
\end{eqnarray}
and for the diagonal elements  (Eq. \ref{deriHlbasisnn}) 
\begin{eqnarray}
\frac{dH_{nn}(l)}{dl} && = 2 \sum_{k \ne n} 
  \frac{ \vert H_{nk}(l) \vert^2 }{H_{nn}(l) - H_{kk}(l) } 
\label{deriHlbasisnnwhite}
\end{eqnarray}
now contain denominators involving differences of diagonal elements
that are usual in perturbation theory.

This direct link with perturbation theory can be clarified as follows.
If one decomposes the generator $\eta$ according to the order 
with respect to off-diagonal elements
\begin{eqnarray}
\eta = \eta^{(1)}+\eta^{(2)}+\eta^{(3)}+...
\label{etadv}
\end{eqnarray}
the dynamics reads
\begin{eqnarray}
\frac{dH(l)}{dl} && = [\eta^{(1)}+\eta^{(2)}+\eta^{(3)}+..., H_{diag}+H_{off}]
\nonumber \\
&& =[\eta^{(1)}, H_{diag}]
+ \left( [\eta^{(1)},H_{off}]+  [\eta^{(2)}, H_{diag}] \right)
+ \left( [\eta^{(2)},H_{off}]+  [\eta^{(3)}, H_{diag}] \right)+..
\label{deriHper}
\end{eqnarray}

White's choice written in Eq. \ref{etawhitenk} for matrix elements
can be translated at the level of operators into the requirement
\begin{eqnarray}
[ \eta^{White} , H_{diag}] = - H_{off}
\label{whitecomm}
\end{eqnarray}
for the first term in the expansion of Eq. \ref{deriHper},
in order to produce the exponential decay for the whole operator $H_{off}$
(i.e. the same exponential decay for all off-diagonal matrix elements
as given by the first term of Eq. \ref{deriHlbasispqwhite}).
The form of Eq. \ref{whitecomm}  to determine 
the generator of unitary transformation corresponds to the Schrieffer-Wolff 
transformations at first order in perturbation theory (see the review \cite{Wolff-Sreview})
 that have been much used recently for random quantum spin chains
to derive renormalization rules for the ground state \cite{refael_fisher}
or for excited states in the Many-Body Localized Phase \cite{rsrgx,rsrgx_bifurcation}.

\subsubsection{ Intermediate sign choice $f^{sgn}(x)  ={\rm sgn}(x)$ }

As an intermediate between the previous Wegner's and White's proposals,
the sign choice \cite{choixsigne}
\begin{eqnarray}
\eta^{sign}_{nk}(l) && = H_{nk}(l) {\rm sgn} (H_{nn}(l) - H_{kk}(l) )
\label{etasgnnk}
\end{eqnarray}
corresponds to the convergence criterion
\begin{eqnarray}
0< - \frac{dI_2^{off}(l)}{dl}  = \frac{dI_2^{diag}(l)}{dl} 
&& = 2 \sum_{n \ne k}  \vert H_{nn}(l)- H_{kk}(l) \vert  \vert H_{nk}(l) \vert^2
\label{I2diagflowcontisign}
\end{eqnarray}
and to the flow equations for the off-diagonal elements
\begin{eqnarray}
\frac{dH_{nq}(l)}{dl} && = - \vert H_{nn}(l)-H_{qq}(l) \vert H_{nq}(l)
+ \sum_{k \ne (n,q)} H_{nk}(l) H_{kq}(l) \big( 
 {\rm sgn} (H_{nn}(l) - H_{kk}(l) ) +  {\rm sgn} (H_{qq}(l) - H_{kk}(l) ) \big)
\label{deriHlbasispqsign}
\end{eqnarray}
and the diagonal elements
\begin{eqnarray}
\frac{dH_{nn}(l)}{dl} && =
2 \sum_{k \ne n}  \vert H_{nk}(l) \vert^2 {\rm sgn} (H_{nn}(l) - H_{kk}(l) ) 
\label{deriHlbasisnnsign}
\end{eqnarray}

\subsubsection{ Comparison of the phase space contraction }

To simplify the discussion,
 let us focus on the case where $H$ is a real symmetric matrix :
there are $M$ diagonal matrix elements $H_{nn}$ with $n=1,..,M$
and $\frac{M^2-M}{2}$ off-diagonal matrix elements $H_{nq}$
 with $1 \leq n<q \leq M$.

Then the dynamical equations for the matrix elements are of the form
\begin{eqnarray}
\frac{dH_{n \leq q}(l)}{dl} && = V_{nq} ( H(l))
\label{defV}
\end{eqnarray}
in terms of the velocity field
\begin{eqnarray}
V_{n<q} ( H) && = - H_{nq}(H_{nn}-H_{qq} ) f(H_{nn}-H_{qq} )
\nonumber \\
&& + \sum_{k \ne (n,q)} 
\left[  \theta_{k<n} H_{kn} +\theta_{n<k} H_{nk}  \right]
\left[  \theta_{k<q} H_{kq} +\theta_{q<k} H_{qk}  \right]
\left( f(H_{nn}-H_{kk})+ f(H_{qq}-H_{kk}) \right)
\label{vnq}
\end{eqnarray}
and
\begin{eqnarray}
V_{nn}(H) && =2 \sum_{k \ne n} \left[ \theta_{k<n} H_{kn}^2 +\theta_{n<k} H_{nk}^2  \right] f(H_{nn}-H_{kk})
\label{vnn}
\end{eqnarray}
where $\theta_{k<n}=1$ if $k<n$ and zero otherwise.

If the initial condition is described by some probability distribution 
$\rho_{l=0}(H)$ with the elementary volume element
\begin{eqnarray}
d {\cal V} \equiv \prod_{n=1}^M dH_{nn}  \prod_{1 \leq n<q\leq M} d{ H}_{ij}
\label{vol}
\end{eqnarray}
the dynamics is governed by the continuity equation
\begin{eqnarray}
\frac{ \partial \rho_{l}(H) }{\partial l} &&  = - \vec \nabla . \left[ \rho_{l}(H) \vec V   \right]
  = -\rho_{l}(H) \left[ \vec \nabla . \vec V   \right] - \vec V .  \vec \nabla \rho_{l}(H) 
\label{continuity}
\end{eqnarray}
where the first term containing the divergence of the velocity field $\left[ \vec \nabla . \vec V   \right] $ represents the contraction of the phase space volume of Eq. \ref{vol}, while the second term contains the advective derivative $\vec V .  \vec \nabla $ familiar from hydrodynamics.

While the off-diagonal directions are always contracting
\begin{eqnarray}
\frac{ \partial V_{n<q} } { \partial H_{n<q} }  
&& =  -   (H_{nn}-H_{qq} ) f(H_{nn}-H_{qq}) <0 
\label{velocityderi}
\end{eqnarray}
the diagonal directions correspond to contraction if $f'(x)<0$
or to expansion if $f'(x)>0$.
\begin{eqnarray}
\frac{ \partial V_{nn} } { \partial H_{nn} } 
&& = 2 \sum_{k \ne n}  H_{kn}^2 f'(H_{nn}-H_{kk})
\label{velocityderifdiag}
\end{eqnarray}

As a consequence, the global resulting divergence of the velocity field
\begin{eqnarray}
 \vec \nabla . \vec V  && = \sum_{1 \leq n \leq M} \frac{ \partial V_{nn} } { \partial H_{nn} }  + \sum_{1 \leq n<q \leq M} \frac{ \partial V_{n<q} } { \partial H_{n<q} }  
\nonumber \\ 
&& = 4 \sum_{1 \leq n<q \leq M} H^2_{nq}  f'(H_{nn}-H_{qq})
- \sum_{1 \leq n<q \leq M}  (H_{nn}-H_{qq} ) f(H_{nn}-H_{qq})
\label{velocitydiv}
\end{eqnarray}
depends on the choice of the function $f$.
For the three cases described above 
\begin{eqnarray}
\frac{d f^{Wegner}}{dx} && = 1 \geq 0
\nonumber \\
\frac{df^{White}}{dx} && = - \frac{1}{x^2} \leq 0
\nonumber \\
\frac{df^{sgn} }{dx}&& = 2 \delta (x)  =0  \ \  \ \  { \rm   for } \ \  \ \  x \ne 0
\label{fvariousderi}
\end{eqnarray}
one obtains the corresponding divergences
\begin{eqnarray}
 \vec \nabla . \vec V^{Wegner}   
&& = 4 \sum_{1 \leq n<q \leq M} H^2_{nq} 
- \sum_{1 \leq n<q \leq M}  (H_{nn}-H_{qq} )^2 
\\
 \vec \nabla . \vec V^{White}  
&& = - 4 \sum_{1 \leq n<q \leq M} \frac{ H^2_{nq} }{ (H_{nn}-H_{qq})^2 }
- \sum_{1 \leq n<q \leq M}  1
\\
 \vec \nabla . \vec V^{sgn}   
&& =  8 \sum_{1 \leq n<q \leq M} H^2_{nq}  \delta(H_{nn}-H_{qq})
- \sum_{1 \leq n<q \leq M}  \vert H_{nn}-H_{qq} \vert
\label{velocitydivres}
\end{eqnarray}
So while the White's flow and the sign flow are always contracting,
Wegner's flow can be expanding in the initial stage as long as the off-diagonal part has not decreased sufficiently.

\subsection{ Toda flow   } 

In the above continuous flows based on the systematic decay of $I_2^{off}$
ensured by generators of the form of Eq. \ref{etaf}, it is impossible to avoid 
the generation of new matrix elements even if the initial condition is sparse.
It is thus interesting to look for other flows that can preserve the sparsity of
the initial matrix.

\subsubsection{ Closed flow for band matrices }

If the initial matrix has some band structure
\begin{eqnarray}
 H_{nk}(l=0) = 0 \ \ {\rm for } \ \ \vert n - k \vert > B 
\label{band}
\end{eqnarray}
it is possible to preserve this structure via the flow if one chooses
the generator
\begin{eqnarray}
\eta^{Toda}_{nk}= H_{nk} {\rm sgn } (k-n )
\label{etatoda}
\end{eqnarray}
because the flow equation for the off-diagonal terms read
\begin{eqnarray}
\frac{dH_{nq}}{dl} && = - (H_{nn}-H_{qq} )  H_{nq} {\rm sgn } (q-n )
+ \sum_{k \ne (n,q)} H_{nk}   H_{kq} \left(  {\rm sgn } (k-n ) + {\rm sgn } (k-q )\right)
\label{todaoff}
\end{eqnarray}
while the diagonal terms evolve according to
\begin{eqnarray}
\frac{dH_{nn}}{dl} && =2 \sum_{k \ne n} \vert H_{nk} \vert^2   {\rm sgn } (k-n )
\label{todadiag}
\end{eqnarray}
The generator of Eq. \ref{etatoda} has been re-discovered 
by Mielke \cite{mielke_band} from the requirement to obtain a closed flow for band matrices, but has actually a long history for the special case of real tridiagonal matrices as we now recall.

\subsubsection{ Toda flow for tridiagonal real matrices }

For the special case of tridiagonal real matrices, 
where the only non-vanishing elements are the diagonal elements $H_{nn}$
and the off-diagonal elements $H_{n,n+1}=H_{n+1,n}$, the closed flow based on
the generator of Eq. \ref{etatoda}
\begin{eqnarray}
\frac{dH_{nn}}{dl} && =2 ( H_{n,n+1}^2- H_{n-1,n}^2 )
\nonumber \\
\frac{dH_{n,n+1}}{dl} && = H_{n,n+1}   (H_{n+1,n+1} -H_{nn} ) 
\label{todatridiag}
\end{eqnarray}
have been much studied 
 under the name of "Toda flow" \cite{henon,flaschka,moser,deift,symes,watkins,chu,elsner,tomei} as a consequence of its relation to the classical integrable Toda lattice \cite{toda} via a change of variables :
 the essential idea is that the flow equation of Eq. \ref{deriHl}
 corresponds to a Lax Pair equation for the integrable Toda model.

\subsubsection{ Convergence towards diagonalization  }

For our present perspective, the most important result
is that this Toda flow converges towards diagonalization 
as first proven by Moser \cite{moser}
\begin{eqnarray}
H_{n,n+1}(l=\infty)=0 
\label{todaoffzero}
\end{eqnarray}
with ordered eigenvalues
\begin{eqnarray}
H_{11}(\infty)>...>H_{NN}(\infty) 
\label{todadiagordered}
\end{eqnarray}

This result can be understood at the level of the differential equations
by rewriting the flow of the diagonal terms 
(Eq. \ref{todatridiag}) as 
\begin{eqnarray}
\frac{dH_{11}}{dl} && =2  H_{1,2}^2 \geq 0
\nonumber \\
\frac{dH_{11}}{dl} +\frac{dH_{22}}{dl} && =2  H_{2,3}^2 \geq 0
\nonumber \\
\frac{dH_{11}}{dl} +\frac{dH_{22}}{dl}+\frac{dH_{33}}{dl} && =2  H_{3,4}^2 \geq 0
\nonumber \\
...
\nonumber \\
\sum_{n=1}^{N-1} \frac{dH_{nn}}{dl} && =2 H_{N-1,N}^2 \geq 0
\label{todatridiagdeb}
\end{eqnarray}
so that $H_{11}(l)$, $(H_{11}(l)+H_{22}(l))$ , etc 
are non-decreasing functions. Since they are bounded as a consequence of the
invariants of the flow, they have to converge
 towards finite values $H_{nn}(l=+\infty)$ at $l=+\infty$.
So the off-diagonal elements $H_{n,n+1}(l)$ representing their derivatives 
(Eq. \ref{todatridiagdeb}) have to vanish $H_{n,n+1}(l=\infty)=0$ at $l=+\infty$. 
The flow of the off-diagonal element $H_{n,n+1}$ (Eq \ref{todatridiag})
can converge towards zero only if the corresponding diagonal elements satisfy
the order $H_{nn}(l) > H_{n+1,n+1}(l) $ asymptotically for large $l$.

\subsubsection{ Interpretation as some continuous limit of the QR-algorithm  }

To understand the physical meaning of the Toda flow, it is important to
stress that it can be interpreted as some continuous limit of the QR-algorithm
recalled in section \ref{sec_qr} as follows.
From the current orthonormal basis $\vert i_l >$ with $i=1,..,N$,
one constructs the $N$ infinitesimally-different vectors
by the application of the Hamiltonian :
\begin{eqnarray}
\vert v_{i_l} > = (1+dl H) \vert i_l > = \vert i_l >(1+ dl < i_l \vert H
\vert i_l >) + \sum_{j \ne i } \vert j_l> (dl < j_l \vert H \vert i_l > ) 
\label{qrinfinitesimal}
\end{eqnarray}
Now one needs to orthonormalize them to obtain a new basis. The Gram-Schmidt procedure
begins with the normalization of the first vector $i=1$.
 Using
\begin{eqnarray}
 <v_{1_l} \vert v_{1_l} > =(1+ 2 dl < i_l \vert H \vert i_l >)
\label{v1scal}
\end{eqnarray}
one obtains the first normalized vector of the new basis as
\begin{eqnarray}
\vert 1_{l+dl} > = \frac{\vert v_{1_l} > }{\sqrt{ <v_{1_l} \vert v_{1_l} >} } 
=  \vert 1_l > + \sum_{ j>1 } \vert j_l> (dl < j_l \vert H
\vert 1_l > ) 
\label{v1norm}
\end{eqnarray}
Then the second vector $\vert v_{2_l} > $ is made orthogonal to the previous vector of Eq. \ref{v1norm} via the computation of the scalar product
\begin{eqnarray}
 <1_{l+dl} \vert  v_{2_l} > = dl ( < 1_l \vert H \vert 2_l > +   < 2_l \vert H \vert 1_l > )
\label{scalarw2t}
\end{eqnarray}
and the construction of
\begin{eqnarray}
\vert w_{2_l} > && = \vert v_{2_l} > - \vert 1_{l+dl} > <1_{l+dl} \vert  v_{2_l} > 
\nonumber \\ && =
 \vert 2_l >(1+ dl < 2_l \vert H
\vert 2_l >) 
- \vert 1_{l} > (dl   < 2_l \vert H \vert 1_l > )
 + \sum_{j > 2 } \vert j_l> (dl < j_l \vert H \vert 2_l > ) 
\label{w2t}
\end{eqnarray}
and its normalization to obtain the second vector of the new basis
\begin{eqnarray}
\vert 2_{l+dl} > = \frac{\vert w_{2_l} > }{\sqrt{ <w_{2_l} \vert w_{2_l} >} } 
=   \vert 2_l >
- \vert 1_{l} > (dl   < 2_l \vert H \vert 1_l > )
 + \sum_{j > 2 } \vert j_l> (dl < j_l \vert H \vert 2_l > ) 
\label{w2norm}
\end{eqnarray}
So one sees that the sign function in the Toda generator directly
comes from this orthonormalization procedure.
Similarly by iteration one obtains all vectors of the new basis as
\begin{eqnarray}
\vert i_{l+dl} > 
=   \vert i_l >
- \sum_{j<i} \vert j_{l} > (dl   < i_l \vert H \vert j_l > )
 + \sum_{j > i } \vert j_l> (dl < j_l \vert H \vert 2_l > ) 
\label{winorm}
\end{eqnarray}
that corresponds exactly to the Toda generator of Eq. \ref{etatoda}.

More discussions on relations between the Toda flow and the
QR algorithm can be found in Refs \cite{deift,symes,watkins,chu,elsner,tomei},
together with the correspondences between
other differential flows and other discrete matrix algorithms.
In particular, one important output of these studies \cite{deift,symes,watkins}
is the formal solution of the Toda flow
 for the running Hamiltonian $H(l)$ in terms of the initial Hamiltonian $H(0)$
\begin{eqnarray}
H(l) = Q^t(l) H(0) Q(l) 
\label{formalsolutoda}
\end{eqnarray}
where the orthogonal matrix $Q(l)$ is the orthogonal matrix appearing in the QR-decomposition
(where $R(l)$ is upper-triangular as in section \ref{sec_qr})
of the operator 
\begin{eqnarray}
e^{ l H(0) } = Q(l) R(l) 
\label{formalsolutoda2}
\end{eqnarray}
To this give a clear physical meaning of the Toda flow.

\section{ Discrete framework for quantum spin Hamiltonians  }

\label{sec_discreteMB}

\subsection{ Expansion of the off-diagonal parts with ladder operators  }

In the expansion of the running Hamiltonian of Eq. \ref{hl}
in terms of Pauli matrices, it is convenient to replace
the off-diagonal Pauli matrices $(\sigma^x,\sigma^y)$ 
 by the linear combinations corresponding
to the ladder operators 
\begin{eqnarray}
\sigma^{+} && =\frac{\sigma^x + i \sigma^y}{2} = 
\begin{pmatrix}
0 & 1 \\
0 & 0 
\end{pmatrix}
\nonumber \\
\sigma^{-} && =\frac{\sigma^x - i \sigma^y}{2} = 
\begin{pmatrix}
0 & 0 \\
1 & 0 
\end{pmatrix}
\label{paulipm}
\end{eqnarray}
because they are nilpotent
\begin{eqnarray}
 (\sigma^+)^2 && = 0
\nonumber \\
  (\sigma^-)^2&& = 0
\label{sigmanil}
\end{eqnarray}
and their products correspond to projectors $\pi^{\sigma^z}$
\begin{eqnarray}
\sigma^+ \sigma^- && = \frac{ 1+\sigma^z }{2} = \begin{pmatrix}
1 & 0 \\
0 & 0 
\end{pmatrix} \equiv \pi^+
\nonumber \\
\sigma^- \sigma^+ && =  \frac{ 1-\sigma^z }{2}  = \begin{pmatrix}
0 & 0 \\
0 & 1 
\end{pmatrix}\equiv \pi^-
\label{spinpmproj}
\end{eqnarray}

The operator containing $q_+$ operators $\sigma^+$ on sites $1 \leq n_1<n_2<..<n_{q_+} \leq N $, $q_-$ operators $\sigma^+$ on different sites $1 \leq m_1<m_2<..<m_{q_-} \leq N $
 and $q_z \geq 0$ operators $\sigma^z$ on further
different sites $1 \leq p_1<p_2<..<p_{q_z} \leq N $
\begin{eqnarray}
X^{\dagger}_{p_1,..p_{q_z};n_1,..,n_{q_+};m_1,..,m_{q_-}} && \equiv 
\sigma_{p_1}^z\sigma_{p_2}^z ... \sigma_{p_{q_z}}^z
\sigma_{n_1}^+\sigma_{n_2}^+ ... \sigma_{n_{q_+}}^+ \
 \sigma_{m_1}^- \sigma_{m_2}^- ... \sigma_{m_{q_-}}^- 
\label{xxqqdagger}
\end{eqnarray}
and its adjoint
\begin{eqnarray}
X_{p_1,..p_{q_z};n_1,..,n_{q_+};m_1,..,m_{q_-}} && =
\sigma_{p_1}^z\sigma_{p_2}^z ... \sigma_{p_{q_z}}^z
\sigma_{n_1}^-\sigma_{n_2}^- ... \sigma_{n_{q_+}}^- 
\ \sigma_{m_1}^+ \sigma_{m_2}^+ ... \sigma_{m_{q_-}}^+ 
\label{xxqq}
\end{eqnarray}
are also nilpotent  ($(X^{\dagger})^2=0=X^2=0$) and are associated to the projectors
 (Eq. \ref{spinpmproj})
\begin{eqnarray}
X^{\dagger}_{p_1,..p_{q_z};n_1,..,n_{q_+};m_1,..,m_{q_-}} X_{p_1,..p_{q_z};n_1,..,n_{q_+};m_1,..,m_{q_-}} && =
\pi_{n_1}^+\pi_{n_2}^+ ... \pi_{n_{q_+}}^+ \ \pi_{m_1}^- \pi_{m_2}^- ... \pi_{m_{q_-}}^- 
\nonumber \\ 
X_{p_1,..p_{q_z};n_1,..,n_{q_+};m_1,..,m_{q_-}} X^{\dagger}_{p_1,..p_{q_z};n_1,..,n_{q_+};m_1,..,m_{q_-}} && =
\pi_{n_1}^-\pi_{n_2}^- ... \pi_{n_{q_+}}^- \ \pi_{m_1}^+ \pi_{m_2}^+ ... \pi_{m_{q_-}}^+ 
\label{xxqqproj}
\end{eqnarray}

The off-diagonal Hamiltonian can be decomposed as a sum over such operators
\begin{eqnarray}
&& H_{off}(l)  = \sum_{ q_+ =1}^{ N }  \sum_{ q_- =0}^{ N-q_+ } \sum_{q_z=0}^{N-q_+-q_-}
 \sum_{ n_{(1 \leq \alpha  \leq q_+)} }
\sum_{ m_{(1 \leq \beta \leq q_-)} } \sum_{ p_{(1 \leq \alpha \leq q_z)}} 
\label{hhoffsum} \\ 
&& \left[ K_{p_1,..p_{q_z};n_1,..,n_{q_+};m_1,..,m_{q_-}}(l)
X^{\dagger}_{p_1,..p_{q_z};n_1,..,n_{q_+},m_1,..,m_{q_-}}
+ K^*_{p_1,..p_{q_z};n_1,..,n_{q_+};m_1,..,m_{q_-}}(l)
X_{p_1,..p_{q_z};n_1,..,n_{q_+},m_1,..,m_{q_-}}\right] 
\nonumber
\end{eqnarray}
where the couplings can be obtained as
\begin{eqnarray}
 K_{p_1,..p_{q_z};n_1,..,n_{q_+};m_1,..,m_{q_-}}(l) = 2^{q_++q_- -N} 
Tr \left( X_{p_1,..p_{q_z};n_1,..,n_{q_+},m_1,..,m_{q_-}} H(l)  \right)
\label{joff}
\end{eqnarray}

\subsection{ Adaptation of the Jacobi algorithm to quantum spin chains } 

The adaptation of the Jacobi diagonalization algorithm for matrices
 (recalled in section \ref{sec_jacobi})
to many-body second-quantized Hamiltonians 
has been introduced by White \cite{white} 
and has been applied recently to the problem
of Many-Body Localization for interacting fermions
 by Rademaker and Ortuno \cite{rademaker}.
In the language of spin chains, the procedure can be summarized as follows.
To suppress a given term 
\begin{eqnarray}
K_{p_1,..p_{q_z};n_1,..,n_{q_+};m_1,..,m_{q_-}}(l)
X^{\dagger}_{p_1,..p_{q_z};n_1,..,n_{q_+},m_1,..,m_{q_-}}
+ K^*_{p_1,..p_{q_z};n_1,..,n_{q_+};m_1,..,m_{q_-}}(l)
X_{p_1,..p_{q_z};n_1,..,n_{q_+},m_1,..,m_{q_-}}
\label{jxdaggerjx}
\end{eqnarray}
 from the off-diagonal Hamiltonian of Eq. \ref{hhoffsum}, 
one needs to consider the generalized unitary rotation of the form \cite{rademaker}
\begin{eqnarray}
 u=e^{\eta}  = && e^{\theta \left( e^{i \phi} X^{\dagger}_{p_1,..p_{q_z};n_1,..,n_{q_+},m_1,..,m_{q_-}} - e^{-i \phi } X_{p_1,..p_{q_z}n_1,..,n_{q_+},m_1,..,m_{q_-}} \right)} 
\nonumber \\  
&& = 1 + (\cos \theta -1) ( X^{\dagger}_{p_1,..p_{q_z};n_1,..,n_{q_+},m_1,..,m_{q_-}}
X_{p_1,..p_{q_z};n_1,..,n_{q_+},m_1,..,m_{q_-}}
\nonumber \\  
&& \ \ \ \ \ \ \ \ \ \ \ \ \ \ \ \ \  \ \ + X_{p_1,..p_{q_z};n_1,..,n_{q_+},m_1,..,m_{q_-}}
X^{\dagger}_{p_1,..p_{q_z};n_1,..,n_{q_+},m_1,..,m_{q_-}}
)
\nonumber \\ 
&& + \sin \theta  ( e^{i \phi }
X^{\dagger}_{p_1,..p_{q_z};n_1,..,n_{q_+},m_1,..,m_{q_-}}
-e^{-i \phi } 
X_{p_1,..p_{q_z};n_1,..,n_{q_+},m_1,..,m_{q_-}}
) 
\label{uxxexpeta}
\end{eqnarray}
where the angles $(\theta,\phi)$ 
have to be chosen to obtain that the transformed coupling of Eq. \ref{joff}
vanishes
\begin{eqnarray}
0&& = K^{new}_{p_1,..p_{q_z};n_1,..,n_{q_+};m_1,..,m_{q_-}} = 2^{q_++q_- -N} 
Tr \left( X_{p_1,..p_{q_z};n_1,..,n_{q_+},m_1,..,m_{q_-}} ( u H u^{\dagger} )  \right)
\nonumber \\ 
&& = 2^{q_++q_- -N} 
Tr \left(  u^{\dagger} X_{p_1,..p_{q_z};n_1,..,n_{q_+},m_1,..,m_{q_-}} u  H   \right)
\label{joffnew}
\end{eqnarray}

Using the transformation of the operator $X$
\begin{eqnarray}
&&  u^{\dagger} X_{p_1,..p_{q_z};n_1,..,n_{q_+},m_1,..,m_{q_-}} u 
 = \cos^2 \theta X_{p_1,..p_{q_z};n_1,..,n_{q_+},m_1,..,m_{q_-}}
 - \sin^2 \theta e^{i 2 \phi} X^{\dagger}_{p_1,..p_{q_z};n_1,..,n_{q_+},m_1,..,m_{q_-}}
\nonumber \\ 
&& \ \ \ \ \ \ \ \ \ \ \ \ \ \ \ \ \  \
- \cos \theta \sin \theta   e^{i \phi }
(
 X^{\dagger}_{p_1,..p_{q_z};n_1,..,n_{q_+},m_1,..,m_{q_-}}X_{p_1,..p_{q_z};n_1,..,n_{q_+},m_1,..,m_{q_-}}
\nonumber \\ 
&& \ \ \ \ \ \ \ \ \ \ \ \ \ \ \ \ \  \  \ \ \ \ \ \ \ \  \ 
-  X_{p_1,..p_{q_z};n_1,..,n_{q_+},m_1,..,m_{q_-}}X^{\dagger}_{p_1,..p_{q_z};n_1,..,n_{q_+},m_1,..,m_{q_-}} )
\label{transfox}
\end{eqnarray}
Eq \ref{joffnew} becomes
\begin{eqnarray}
0&& = K^{new}_{p_1,..p_{q_z};n_1,..,n_{q_+};m_1,..,m_{q_-}} 
\nonumber \\ 
&& = \cos^2 \theta K_{p_1,..p_{q_z};n_1,..,n_{q_+},m_1,..,m_{q_-}}  
 - \sin^2 \theta e^{i 2 \phi} K^{*}_{p_1,..p_{q_z};n_1,..,n_{q_+},m_1,..,m_{q_-}} 
\nonumber \\ 
&& - \cos \theta \sin \theta   e^{i \phi }
2^{q_++q_- -N} Tr \left(
\left[\pi_{n_1}^+ ... \pi_{n_{q_+}}^+ \ \pi_{m_1}^-  ... \pi_{m_{q_-}}^- 
-
\pi_{n_1}^- ... \pi_{n_{q_+}}^- \ \pi_{m_1}^+ ... \pi_{m_{q_-}}^+ 
 \right]   H_{diag}   \right)
\label{joffnewres}
\end{eqnarray}

So the angle $\phi$ has to be chosen as the phase of the coupling
(analog to Eq. \ref{phil})
\begin{eqnarray}
e^{i \phi} && = \frac{ K_{p_1,..p_{q_z};n_1,..,n_{q_+};m_1,..,m_{q_-}} }
{\vert  K_{p_1,..p_{q_z};n_1,..,n_{q_+};m_1,..,m_{q_-}}   \vert}
\label{philspin}
\end{eqnarray}
and the angle $\theta$ has to be chosen as (analog to Eq. \ref{etaulin})
\begin{eqnarray}
 \tan (2 \theta )  = \frac{ 2 \vert K_{p_1,..p_{q_z};n_1,..,n_{q_+},m_1,..,m_{q_-}} \vert }
{2^{q_++q_- -N} Tr \left(
\left[\pi_{n_1}^+ ... \pi_{n_{q_+}}^+ \ \pi_{m_1}^-  ... \pi_{m_{q_-}}^- 
-
\pi_{n_1}^- ... \pi_{n_{q_+}}^- \ \pi_{m_1}^+ ... \pi_{m_{q_-}}^+ 
 \right]   H_{diag}   \right)} 
\label{joffnewfinal}
\end{eqnarray}
where the denominator only involves the $z-$couplings 
concerning the spins $n_{\alpha},m_{\beta}$.

As an example, the XXZ chain with random fields of Eq. \ref{XXZh}
contains initially  off-diagonal terms of the form
\begin{eqnarray}
 J_i  ( \sigma_i^+ \sigma_{i+1}^-+\sigma_i^- \sigma_{i+1}^+ )
= K_{0;i;i+1}(l) \left(X^{\dagger}_{0;i,i+1} +X_{0;i,i+1}  \right)
\label{jxdaggerjxex}
\end{eqnarray}
This given term can be suppressed via the unitary transformation
with $\phi=0$
\begin{eqnarray}
&& u=e^{\eta}  = e^{\theta \left( \sigma_i^+ \sigma_{i+1}^-- \sigma_i^- \sigma_{i+1}^+\right)} 
\nonumber \\  
&& = 1 + (\cos \theta -1) ( \pi^+_i \pi_{i+1}^- +  \pi^-_i \pi_{i+1}^+) 
 + \sin \theta  ( \sigma_i^+ \sigma_{i+1}^-- \sigma_i^- \sigma_{i+1}^+  ) 
\nonumber \\  
&& = 1 + (\cos \theta -1) \frac{1-\sigma_i^z \sigma_{i+1}^z}{2}  
 + \sin \theta  ( \sigma_i^+ \sigma_{i+1}^-- \sigma_i^- \sigma_{i+1}^+  ) 
\label{uxxzexpeta}
\end{eqnarray}
where the angle $\theta$ has to be chosen as (Eq. \ref{joffnewfinal})
\begin{eqnarray}
 \tan (2 \theta )  && = \frac{ 2 J_i }
{2^{2 -N} Tr \left(
\left[\pi_{i}^+ \pi_{i+1}^- -\pi_{i}^- \pi_{i+1}^+  \right]   H_{diag}   \right)} 
= \frac{  J_i }
{2^{ -N} Tr \left( (\sigma_i^z - \sigma_{i+1}^z)   H_{diag}   \right)} 
\nonumber \\  
&& = \frac{  J_i }
{ h_i- h_{i+1}} 
\label{joffnewfinalxxz}
\end{eqnarray}

As in the Jacobi algorithm, this method tends to generate all possible
off-diagonal couplings in the running Hamiltonian of Eq. \ref{hhoffsum}
even if one starts from a sparse initial condition,
so that one needs to introduce some truncations in the numerical application
of this procedure : we refer to References \cite{white,rademaker}
for discussions and examples of numerical results that can be obtained.

\section{ Continuous framework for quantum spin Hamiltonians  } 

\label{sec_continuousMB}

In the continuous framework, the most general anti-hermitian
generator can be expanded in terms of all the operators involved
 in the off-diagonal part of Eq. \ref{hhoffsum}
\begin{eqnarray}
&& \eta(l)  = \sum_{ q_+ =1}^{ N }  \sum_{ q_- =0}^{ N-q_+ } \sum_{q_z=0}^{N-q_+-q_-}
 \sum_{ n_{(1 \leq \alpha  \leq q_+)} }
\sum_{ m_{(1 \leq \beta \leq q_-)} } \sum_{ p_{(1 \leq \alpha \leq q_z)}} 
\theta_{p_1,..p_{q_z};n_1,..,n_{q_+};m_1,..,m_{q_-}}(l)
\label{etahoffsum} \\ 
&& \left[  e^{ i \phi_{p_1,..p_{q_z};n_1,..,n_{q_+};m_1,..,m_{q_-}}(l) }
X^{\dagger}_{p_1,..p_{q_z};n_1,..,n_{q_+},m_1,..,m_{q_-}}
-  e^{ - i \phi_{p_1,..p_{q_z};n_1,..,n_{q_+};m_1,..,m_{q_-}}(l) }
X_{p_1,..p_{q_z};n_1,..,n_{q_+},m_1,..,m_{q_-}}\right] 
\nonumber
\end{eqnarray}
where the generalized angles $\theta_{p_1,..p_{q_z};n_1,..,n_{q_+};m_1,..,m_{q_-}}(l) $ and $\phi_{p_1,..p_{q_z};n_1,..,n_{q_+};m_1,..,m_{q_-}}(l) $ 
define the choice of $\eta$. 

In particular, the Wegner's choice of Eq. \ref{etawegner} 
\begin{eqnarray}
\eta^{Wegner} (l) = [ H_{diag}(l), H(l) ] =  [ H_{diag}(l), H_{off}(l) ] 
\label{wegners}
\end{eqnarray}
 has been applied to many different condensed matter problems (see the reviews \cite{wegner_reviews}, the book \cite{book_kehrein} and references therein).

The White's choice of Eq. \ref{whitecomm}
 \begin{eqnarray}
[ \eta^{White}(l) , H_{diag}(l)] = - H_{off}(l)
\label{whitecomm2}
\end{eqnarray}
has been applied numerically and compared to the discrete Jacobi framework in Ref \cite{white}. 

As in the Jacobi method, 
 these Wegner's and White's flows based on the systematic decay of $I_2^{off}$
tend to generate all possible
off-diagonal couplings in the running Hamiltonian of Eq. \ref{hhoffsum},
and one has again to introduce some truncation in the numerical implementation.

\section{ Adaptation of the idea of the Toda flow to quantum spin chains }

\label{sec_todaspin}

\subsection{ Ansatz for a simplified closed flow }

Let us now focus on the XXZ chain of Eq. \ref{XXZh} as the initial state.
We would like to define the 'best closed flow'
of the form
\begin{eqnarray}
H^{XXZ}(l) && = H_{diag}\{\sigma_r^z\} + \sum_n
{\cal J}_{n}\{\sigma_{r \ne (n,n+1)}^z\}  ( \sigma_n^+ \sigma_{n+1}^- +\sigma_n^- \sigma_{n+1}^+ )
\label{htoda}
\end{eqnarray}
where the diagonal part is written as $H_{diag}\{\sigma_r^z\} $ 
to emphasize that it depends on the $N$ operators $\sigma_r^z $ via the expansion
involving $2^N$ real running coefficients $H_{a_1...a_N }(l)  $ 
\begin{eqnarray}
H_{diag}\{\sigma_r^z\} =  \sum_{a_1=0,z} ...  \sum_{a_N=0,z}
H_{a_1...a_N } (l) \sigma_1^{(a_1)}  \sigma_2^{(a_2)}  ...  \sigma_N^{(a_N)} 
\label{hldiagr}
\end{eqnarray}
as in the final state at $l=\infty$ of Eq. \ref{hldiagxxzinfty}.
Similarly, the prefactor of the elementary off-diagonal 
operator $( \sigma_n^+ \sigma_{n+1}^- +\sigma_n^- \sigma_{n+1}^+ )$ in Eq. \ref{htoda}
is written as
${\cal J}_{n}\{\sigma_{r \ne (n,n+1)}^z\} $
to emphasize that it depends on the $(N-2)$ operators $\sigma_{r \ne (n,n+1)}^z $ via the expansion
involving $2^{N-2}$ real running coefficients $ J_n(a_1,..,a_{n-1},a_{n+2}..,a_N;l) $ 
\begin{eqnarray}
{\cal J}_{n}\{\sigma_{r \ne (n,n+1)}^z\} =   \sum_{a_1=0,z}.. .\sum_{a_{n-1}=0,z}\sum_{a_{n+2}=0,z} ...  \sum_{a_N=0,z}
J_n(a_1,..,a_{n-1},a_{n+2}..,a_N;l)   \sigma_1^{(a_1)} ... \sigma_{n-1}^{(a_{n-1})} \sigma_{n+2}^{(a_{n+2})} ...  \sigma_N^{(a_N)} 
\label{jtoda}
\end{eqnarray}
So the simplification with respect to the most general off-diagonal Hamiltonian
of Eq. \ref{hhoffsum} is that only terms corresponding to the operators 
$X_{0,n,n+1}= \sigma_n^+ \sigma_{n+1}^-$ are included.

Accordingly, the generator is chosen to include only 
the off-diagonal operators $X_{0,n,n+1}= \sigma_n^+ \sigma_{n+1}^- $ (instead of the most general form of Eq. \ref{etahoffsum})
\begin{eqnarray}
\eta && =\sum_{i}
 \Theta_{i}\{\sigma_{k \ne (i,i+1)} ^z\}  ( \sigma_i^+ \sigma_{i+1}^- - \sigma_i^- \sigma_{i+1}^+ )
\label{etatheta}
\end{eqnarray}
where 
$\Theta_{i}\{\sigma_{k \ne (i,i+1)}^z\}  $
 depends on the $(N-2)$ operators $\sigma_{k \ne (i,i+1)}^z$ via the expansion
involving $2^{N-2}$ real running coefficients 
$ \theta_i(a_1,..,a_{i-1},a_{i+2}..,a_N;l) $ 
\begin{eqnarray}
\Theta_{i}\{\sigma_{r \ne (i,i+1)}^z\} =   \sum_{a_1=0,z}.. .\sum_{a_{i-1}=0,z}\sum_{a_{i+1}=0,z} ...  \sum_{a_N=0,z}
\theta_i(a_1,..,a_{i-1},a_{i+2}..,a_N;l)   \sigma_1^{(a_1)} ... \sigma_{i-1}^{(a_{i-1})} \sigma_{i+2}^{(a_{i+2})} ...  \sigma_N^{(a_N)} 
\label{thetaeta}
\end{eqnarray}
Our goal is to choose the generator $\eta$ to maintain the flow
of the Hamiltonian of Eq. \ref{htoda}
as 'closed' as possible.

\subsection{ Flow equation  }

With the above Ansatz, the flow equation reads
\begin{eqnarray}
\frac{dH}{dl}  = [ \eta, H] 
&& = \sum_{i}
[ \Theta_{i}\{\sigma_{k \ne (i,i+1)} ^z\}  ( \sigma_i^+ \sigma_{i+1}^- - \sigma_i^- \sigma_{i+1}^+ ) ,  H_{diag}\{\sigma_r^z\}]
\nonumber \\ && 
+ \sum_{i}
[ \Theta_{i}\{\sigma_{k \ne (i,i+1)} ^z\}  ( \sigma_i^+ \sigma_{i+1}^- - \sigma_i^- \sigma_{i+1}^+ ) , 
{\cal J}_{i}\{\sigma_{r \ne (i,i+1)}^z\}  ( \sigma_i^+ \sigma_{i+1}^- +\sigma_i^- \sigma_{i+1}^+ )]
\nonumber \\ && 
+ \sum_{i}
[ \Theta_{i}\{\sigma_{k \ne (i,i+1)} ^z\}  ( \sigma_i^+ \sigma_{i+1}^- - \sigma_i^- \sigma_{i+1}^+ ) , 
{\cal J}_{i-1}\{\sigma_{r \ne (i-1,i)}^z\}  ( \sigma_{i-1}^+ \sigma_{i}^- +\sigma_{i-1}^- \sigma_{i}^+ )]
\nonumber \\ && 
+ \sum_{i}
[ \Theta_{i}\{\sigma_{k \ne (i,i+1)} ^z\}  ( \sigma_i^+ \sigma_{i+1}^- - \sigma_i^- \sigma_{i+1}^+ ) ,  
{\cal J}_{i+1}\{\sigma_{r \ne (i+1,i+2)}^z\}  ( \sigma_{i+1}^+ \sigma_{i+2}^- +\sigma_{i+1}^- \sigma_{i+2}^+ )]
\nonumber \\ && 
+ \sum_{i}
[ \Theta_{i}\{\sigma_{k \ne (i,i+1)} ^z\}  ( \sigma_i^+ \sigma_{i+1}^- - \sigma_i^- \sigma_{i+1}^+ ) ,  \sum_{ n \ne (i-1,i,i+1)}
{\cal J}_{n}\{\sigma_{r \ne (n,n+1)}^z\}  ( \sigma_n^+ \sigma_{n+1}^- +\sigma_n^- \sigma_{n+1}^+ )]
\label{flowtoda}
\end{eqnarray}

To compute these various commutators,
it is useful to introduce for any function $f \{\sigma_r^z\} $
its expansion with respect to a given spin $\sigma_i^z$
\begin{eqnarray}
f \{\sigma_r^z\} && =f^{(i,0)} \{\sigma_{r \ne i}^z\} 
 + \sigma_{i}^z f^{(i,1)} \{\sigma_{r \ne i}^z\} 
\label{dv1spinz}
\end{eqnarray}
where the two auxiliary functions can be obtained by some
traces with respect to the single spin $\sigma_i $
\begin{eqnarray}
f^{(i,0)} \{\sigma_{r \ne i}^z\} && = \frac{1}{2} Tr_{\sigma_i} (f \{\sigma_r^z\}  )
\nonumber \\ 
f^{(i,1)} \{\sigma_{r \ne i}^z\} && =\frac{1}{2} Tr_{\sigma_i} (\sigma_{i}^z f \{\sigma_r^z\}  )
\label{dv1spinzaux}
\end{eqnarray}
Similarly, its expansion with respect to a pair of
 given spins $(\sigma_i^z ,  \sigma_j^z )$ reads
\begin{eqnarray}
f \{\sigma_r^z\} && =f^{(i,0)(j,0)} \{\sigma_{r \ne i,j}^z\} 
 + \sigma_{i}^z f^{(i,1)(j,0)} \{\sigma_{r \ne i,j}^z\} 
+ \sigma_{j}^z f^{(i,0)(j,1)} \{\sigma_{r \ne i,j}^z\} 
+ \sigma_{i}^z \sigma_{j}^z f^{(i,1)(j,1)} \{\sigma_{r \ne i,j}^z\} 
\label{dv2spinz}
\end{eqnarray}
where the four auxiliary functions can be obtained by some
traces with respect to the two spins $(\sigma_i,\sigma_j) $
\begin{eqnarray}
f^{(i,0)(j,0)} \{\sigma_{r \ne i,j}^z\} && = \frac{1}{2^2} Tr_{\sigma_i,\sigma_j} (f \{\sigma_r^z\}  )
\nonumber \\ 
f^{(i,1)(j,0)} \{\sigma_{r \ne i,j}^z\} && =\frac{1}{2^2} Tr_{\sigma_i,\sigma_j} (\sigma_{i}^z f \{\sigma_r^z\}  ) 
\nonumber \\ 
f^{(i,0)(j,1)} \{\sigma_{r \ne i,j}^z\} && =\frac{1}{2^2} Tr_{\sigma_i,\sigma_j} (\sigma_{j}^z f \{\sigma_r^z\}  ) 
\nonumber \\ 
f^{(i,1)(j,1)} \{\sigma_{r \ne i,j}^z\} && =\frac{1}{2^2} Tr_{\sigma_i,\sigma_j} (\sigma_{i}^z \sigma_{j}^z f \{\sigma_r^z\}  ) 
\label{dv2spinzaux}
\end{eqnarray}

Using the expansion of Eq. \ref{dv2spinz} for the diagonal Hamiltonian
\begin{eqnarray}
H_{diag} \{\sigma_r^z\} && =H_{diag}^{(i,0)(i+1,0)} \{\sigma_{r \ne i,i+1}^z\} 
 + \sigma_{i}^z H_{diag}^{(i,1)(i+1,0)} \{\sigma_{r \ne i,i+1}^z\} 
+ \sigma_{i+1}^z H_{diag}^{(i,0)(i+1,1)} \{\sigma_{r \ne i,i+1}^z\} 
\nonumber \\ 
&& + \sigma_{i}^z \sigma_{i+1}^z H_{diag}^{(i,1)(i+1,1)} \{\sigma_{r \ne i,i+1}^z\} 
\label{dv2spinshdiag}
\end{eqnarray}
the first line of the flow Eq. \ref{flowtoda} becomes
\begin{eqnarray}
\left( \frac{dH}{dl}\right)_{first} 
 = 2 \sum_{i}\Theta_{i}\{\sigma_{k \ne (i,i+1)} ^z\}
\left(  H_{diag}^{(i,0)(i+1,1)} \{\sigma_{r \ne i,i+1}^z\}
- H_{diag}^{(i,1)(i+1,0)} \{\sigma_{r \ne i,i+1}^z\}
\right)
( \sigma_i^+ \sigma_{i+1}^-   + \sigma_i^- \sigma_{i+1}^+ )
\label{flowtoda1}
\end{eqnarray}
It is thus compatible with the flow Ansatz of Eq. \ref{htoda}
and will actually be the only contribution to the flow of
the ${\cal J}_{i}\{\sigma_{r \ne (n,n+1)}^z\} $ that reads
\begin{eqnarray}
\frac{ d {\cal J}_{i}\{\sigma_{r \ne (i,i+1)}^z\} }{dl}
= 2 \Theta_{i}\{\sigma_{k \ne (i,i+1)} ^z\}
\left(  H_{diag}^{(i,0)(i+1,1)} \{\sigma_{r \ne i,i+1}^z\}
- H_{diag}^{(i,1)(i+1,0)} \{\sigma_{r \ne i,i+1}^z\}
\right)
\label{flotjtoda}
\end{eqnarray}

The second line of Eq. \ref{flowtoda}
\begin{eqnarray}
\left(\frac{dH}{dl}\right)_{second}  
&& \equiv \sum_{i}
[ \Theta_{i}\{\sigma_{k \ne (i,i+1)} ^z\}  ( \sigma_i^+ \sigma_{i+1}^- - \sigma_i^- \sigma_{i+1}^+ ) , 
{\cal J}_{i}\{\sigma_{r \ne (i,i+1)}^z\}  ( \sigma_i^+ \sigma_{i+1}^- +\sigma_i^- \sigma_{i+1}^+ )] 
\nonumber \\ && = \sum_{i}\Theta_{i}\{\sigma_{k \ne (i,i+1)} ^z\} {\cal J}_{i}\{\sigma_{r \ne (i,i+1)}^z\}
(\sigma_i^z - \sigma_{i+1}^z )
\label{flowtoda2}
\end{eqnarray}
is also compatible with the flow Ansatz of Eq. \ref{htoda}
and will actually be the only contribution to the flow of
the diagonal part that reads
\begin{eqnarray}
\frac{ d H_{diag}\{\sigma_r^z\}  }{dl} =  \sum_{i}\Theta_{i}\{\sigma_{k \ne (i,i+1)} ^z\} {\cal J}_{i}\{\sigma_{r \ne (i,i+1)}^z\}
(\sigma_i^z - \sigma_{i+1}^z )
\label{flothdiagtoda}
\end{eqnarray}

Using the notations of Eq. \ref{dv1spinz},
the third and the fourth lines of Eq. \ref{flowtoda}
lead to the global result
\begin{eqnarray}
&&  \left( \frac{dH}{dl} \right)_{third} + \left(  \frac{dH}{dl} \right)_{fourth}  
 \equiv \sum_{i}
[ \Theta_{i}\{\sigma_{k \ne (i,i+1)} ^z\}  ( \sigma_i^+ \sigma_{i+1}^- - \sigma_i^- \sigma_{i+1}^+ ) , 
{\cal J}_{i-1}\{\sigma_{r \ne (i-1,i)}^z\}  ( \sigma_{i-1}^+ \sigma_{i}^- +\sigma_{i-1}^- \sigma_{i}^+ )]
\nonumber \\ 
&& \ \ \ \ \ \ \ \ \ \ \ \ \ \ \ \ \ \ \ \ \ \ \ \ \ \ \ \ + \sum_{i}
[ \Theta_{i}\{\sigma_{k \ne (i,i+1)} ^z\}  ( \sigma_i^+ \sigma_{i+1}^- - \sigma_i^- \sigma_{i+1}^+ ) ,  
{\cal J}_{i+1}\{\sigma_{r \ne (i+1,i+2)}^z\}  ( \sigma_{i+1}^+ \sigma_{i+2}^- +\sigma_{i+1}^- \sigma_{i+2}^+ )]
\nonumber \\ 
&&  =
\sum_{i}( \Theta_{i}^{(i-1,0)} \{\sigma_{k \ne (i-1,i,i+1)} ^z\}{\cal J}^{(i+1,0)}_{i-1}\{\sigma_{r \ne (i-1,i,i+1)}^z\} 
+\Theta_{i}^{(i-1,1)} \{\sigma_{k \ne (i-1,i,i+1)} ^z\} {\cal J}^{(i+1,1)}_{i-1}\{\sigma_{r \ne (i-1,i,i+1)}^z\} 
\nonumber \\ 
&& -{\cal J}^{(i-1,0)}_{i}\{\sigma_{r \ne (i-1,i,i+1)}^z\} 
 \Theta_{i-1}^{(i+1,0)} \{\sigma_{k \ne (i-1,i,i+1)} ^z\}
-{\cal J}^{(i-1,1)}_{i}\{\sigma_{r \ne (i-1,i,i+1)}^z\}
\Theta_{i-1}^{(i+1,1)} \{\sigma_{k \ne (i-1,i,i+1)} ^z\}
)
\nonumber \\ 
&& (\sigma_{i-1}^+ \sigma_i^z \sigma_{i+1}^-
+ \sigma_{i-1}^- \sigma_{i}^z  \sigma_{i+1}^+ )
\nonumber \\ 
&& +
\sum_{i} ( {\cal J}^{(i+1,0)}_{i-1}\{\sigma_{r \ne (i-1,i,i+1)}^z\}
\Theta_{i}^{(i-1,1)} \{\sigma_{k \ne (i-1,i,i+1)} ^z\}
+ {\cal J}^{(i+1,1)}_{i-1}\{\sigma_{r \ne (i-1,i,i+1)}^z\} 
\Theta_{i}^{(i-1,0)} \{\sigma_{k \ne (i-1,i,i+1)} ^z\}
\nonumber \\ 
&& - \Theta_{i-1}^{(i+1,0)} \{\sigma_{k \ne (i-1,i,i+1)} ^z\}
{\cal J}^{(i-1,1)}_{i}\{\sigma_{r \ne (i-1,i,i+1)}^z\}
- \Theta_{i-1}^{(i+1,1)} \{\sigma_{k \ne (i-1,i,i+1)} ^z\}
{\cal J}^{(i-1,0)}_{i}\{\sigma_{r \ne (i-1,i,i+1)}^z\}
)
\nonumber \\ 
&& (  \sigma_{i-1}^+ \sigma_{i+1}^- + \sigma_{i-1}^-  \sigma_{i+1}^+ )
\label{flowtoda3et4}
\end{eqnarray}
So these terms tend to generate off-diagonal terms
containing two ladder operators concerning two spins at distance two
of the form $ (  \sigma_{i-1}^+ \sigma_{i+1}^- + \sigma_{i-1}^-  \sigma_{i+1}^+  )$ 
that are not present
in the Ansatz of Eq. \ref{htoda} : to avoid the creation of these new terms,
one can require that the prefactors containing $\sigma^z$ operators identically
vanish and one obtains the constraints
\begin{eqnarray}
&&  \Theta_{i}^{(i-1,0)} \{\sigma_{k \ne (i-1,i,i+1)} ^z\}{\cal J}^{(i+1,0)}_{i-1}\{\sigma_{r \ne (i-1,i,i+1)}^z\} 
+\Theta_{i}^{(i-1,1)} \{\sigma_{k \ne (i-1,i,i+1)} ^z\} {\cal J}^{(i+1,1)}_{i-1}\{\sigma_{r \ne (i-1,i,i+1)}^z\} 
\nonumber \\ 
&& = {\cal J}^{(i-1,0)}_{i}\{\sigma_{r \ne (i-1,i,i+1)}^z\} 
 \Theta_{i-1}^{(i+1,0)} \{\sigma_{k \ne (i-1,i,i+1)} ^z\}
+ {\cal J}^{(i-1,1)}_{i}\{\sigma_{r \ne (i-1,i,i+1)}^z\}
\Theta_{i-1}^{(i+1,1)} \{\sigma_{k \ne (i-1,i,i+1)} ^z\}
)
\label{flowtoda3et4vanish1}
\end{eqnarray}
and
\begin{eqnarray}
&& {\cal J}^{(i+1,0)}_{i-1}\{\sigma_{r \ne (i-1,i,i+1)}^z\}
\Theta_{i}^{(i-1,1)} \{\sigma_{k \ne (i-1,i,i+1)} ^z\}
+ {\cal J}^{(i+1,1)}_{i-1}\{\sigma_{r \ne (i-1,i,i+1)}^z\} 
\Theta_{i}^{(i-1,0)} \{\sigma_{k \ne (i-1,i,i+1)} ^z\}
\nonumber \\ 
&& = \Theta_{i-1}^{(i+1,0)} \{\sigma_{k \ne (i-1,i,i+1)} ^z\}
{\cal J}^{(i-1,1)}_{i}\{\sigma_{r \ne (i-1,i,i+1)}^z\}
+ \Theta_{i-1}^{(i+1,1)} \{\sigma_{k \ne (i-1,i,i+1)} ^z\}
{\cal J}^{(i-1,0)}_{i}\{\sigma_{r \ne (i-1,i,i+1)}^z\}
\label{flowtoda3et4vanish2}
\end{eqnarray}
It is clear that a simple solution for the functions $\Theta_{i} \{\sigma_{k \ne (i,i+1)} ^z\} $
defining the generator $\eta$ is the choice
that generalizes the Toda case of Eq \ref{etatoda}
\begin{eqnarray}
\Theta^{Toda}_{i} \{\sigma_{k \ne (i,i+1)} ^z\} = {\cal J}_{i} \{\sigma_{k \ne (i,i+1)} ^z\}
\label{todachoicespin}
\end{eqnarray}
Then the contribution of Eq. \ref{flowtoda3et4} vanish
\begin{eqnarray}
&&  \left( \frac{dH}{dl} \right)^{Toda}_{third} + \left(  \frac{dH}{dl} \right)^{Toda}_{fourth} =0
\label{todachoicespinvan}
\end{eqnarray}

Finally, using the notations of Eq. \ref{dv2spinz}, the fifth and last line
of Eq. \ref{flowtoda}
\begin{eqnarray}
&& \left( \frac{dH}{dl} \right)_{fifth}  
 \equiv \sum_{i}
[ \Theta_{i}\{\sigma_{k \ne (i,i+1)} ^z\}  ( \sigma_i^+ \sigma_{i+1}^- - \sigma_i^- \sigma_{i+1}^+ ) ,  \sum_{ n \ne (i-1,i,i+1)}
{\cal J}_{n}\{\sigma_{r \ne (n,n+1)}^z\}  ( \sigma_n^+ \sigma_{n+1}^- +\sigma_n^- \sigma_{n+1}^+ )]
\nonumber \\
&&= 2  \sum_{i}\sum_{ n \ne (i-1,i,i+1)}
 \left( \Theta_{i}^{(n,0)(n+1,0)}\{\sigma_{k \ne (i,i+1,n,n+1)} ^z\}
-  \Theta_{i}^{(n,1)(n+1,1)}\{\sigma_{k \ne (i,i+1,n,n+1)} ^z\} \right) 
\nonumber \\
&& \left({\cal J}_{n}^{(i,0)(i+1,1)}\{\sigma_{r \ne (i,i+1,n,n+1)}^z\} - {\cal J}_{n}^{(i,1)(i+1,0)}\{\sigma_{r \ne (i,i+1,n,n+1)}^z\}\right)
( \sigma_i^+ \sigma_{i+1}^- + \sigma_i^- \sigma_{i+1}^+  )
 ( \sigma_n^+ \sigma_{n+1}^- +\sigma_n^- \sigma_{n+1}^+ )
\nonumber \\
&& + 2 \sum_{i} \sum_{ n \ne (i-1,i,i+1)}
\left(  \Theta_{i}^{(n,1)(n+1,0)}\{\sigma_{k \ne (i,i+1,n,n+1)} ^z\}
- \Theta_{i}^{(n,0)(n+1,1)}\{\sigma_{k \ne (i,i+1,n,n+1)} ^z\}\right)
\nonumber \\
&& \left({\cal J}_{n}^{(i,0)(i+1,0)}\{\sigma_{r \ne (i,i+1,n,n+1)}^z\}
-  {\cal J}_{n}^{(i,1)(i+1,1)}\{\sigma_{r \ne (i,i+1,n,n+1)}^z\}\right)
( \sigma_i^+ \sigma_{i+1}^- - \sigma_i^- \sigma_{i+1}^+ )
(  \sigma_n^+ \sigma_{n+1}^- - \sigma_n^- \sigma_{n+1}^+ )
\label{flowtoda5ter}
\end{eqnarray}
tends to generate off-diagonal terms
containing four ladder operators concerning two pairs of consecutive spins
$(i,i+1)$ and $(n,n+1)$ 
that are not present
in the Ansatz of Eq. \ref{htoda}.
With the choice of Eq. \ref{todachoicespin}, Eq. \ref{flowtoda5ter}
can be reduced to 
\begin{eqnarray}
&& \left( \frac{dH}{dl} \right)_{fifth}  
\nonumber \\
&&= 4  \sum_{i}\sum_{ n \ne (i-1,i,i+1)}
 \left( {\cal J}_{i}^{(n,0)(n+1,0)}\{\sigma_{k \ne (i,i+1,n,n+1)} ^z\}
-  {\cal J}_{i}^{(n,1)(n+1,1)}\{\sigma_{k \ne (i,i+1,n,n+1)} ^z\} \right) 
\nonumber \\
&& \left({\cal J}_{n}^{(i,0)(i+1,1)}\{\sigma_{r \ne (i,i+1,n,n+1)}^z\} - {\cal J}_{n}^{(i,1)(i+1,0)}\{\sigma_{r \ne (i,i+1,n,n+1)}^z\}\right)
 ( \sigma_i^+ \sigma_{i+1}^- \sigma_n^- \sigma_{n+1}^+ 
  + \sigma_i^- \sigma_{i+1}^+\sigma_n^+ \sigma_{n+1}^- )
\label{flowtoda5toda}
\end{eqnarray}

\subsection { Definition of the analog of the Toda flow }

In summary, we have explained in the previous section that the choice
of Eq. \ref{todachoicespin}
\begin{eqnarray}
\Theta^{Toda}_{i} \{\sigma_{k \ne (i,i+1)} ^z\} = {\cal J}_{i} \{\sigma_{k \ne (i,i+1)} ^z\}
\label{todachoicespin2}
\end{eqnarray}
for the generator of Eq. \ref{etatoda} corresponds to
 the best approximation of a closed flow for the running Hamiltonian of Eq. \ref{htoda} because it suppresses the generation of the most important new off-diagonal
containing two ladder operators concerning non-consecutive spins.
The only non-closed term involves four ladder operators (Eq. \ref{flowtoda5toda})
\begin{eqnarray}
&& \left( \frac{dH}{dl} \right)_{non-closed}  
= 4  \sum_{i}\sum_{ n \ne (i-1,i,i+1)}
 \left( {\cal J}_{i}^{(n,0)(n+1,0)}\{\sigma_{k \ne (i,i+1,n,n+1)} ^z\}
-  {\cal J}_{i}^{(n,1)(n+1,1)}\{\sigma_{k \ne (i,i+1,n,n+1)} ^z\} \right) 
\nonumber \\
&& \left({\cal J}_{n}^{(i,0)(i+1,1)}\{\sigma_{r \ne (i,i+1,n,n+1)}^z\} - {\cal J}_{n}^{(i,1)(i+1,0)}\{\sigma_{r \ne (i,i+1,n,n+1)}^z\}\right)
 ( \sigma_i^+ \sigma_{i+1}^- \sigma_n^- \sigma_{n+1}^+ 
  + \sigma_i^- \sigma_{i+1}^+\sigma_n^+ \sigma_{n+1}^- )
\label{flowtodanonclosed}
\end{eqnarray}

With the choice of Eq. \ref{todachoicespin2},
the coupled flow equations for the diagonal part $H_{diag}\{ \sigma^z_k \} $
and for the generalized couplings ${\cal J}_{i} \{\sigma_{k \ne (i,i+1)} ^z\} $
of Eqs \ref{flotjtoda} and \ref{flothdiagtoda}
become
\begin{eqnarray}
\frac{dH_{diag}\{ \sigma^z_k \}}{dl}
&&  = \sum_{i}  {\cal J}_{i}^2\{\sigma_{r \ne (i,i+1)}^z\}
(\sigma_i^z - \sigma_{i+1}^z )
\label{flowtodahdiag}
\end{eqnarray}
and
\begin{eqnarray}
 \frac{d {\cal J}_{i} \{\sigma_{k \ne (i,i+1)} ^z\}}{dl}
&&  = 2 {\cal J}_{i}\{\sigma_{k \ne (i,i+1)} ^z\}
\left(  H_{diag}^{(i,0)(i+1,1)} \{\sigma_{r \ne i,i+1}^z\}
- H_{diag}^{(i,1)(i+1,0)} \{\sigma_{r \ne i,i+1}^z\}
\right)
\label{flowtodajiz}
\end{eqnarray}
with the notations of Eq. \ref{dv2spinzaux} and Eq. \ref{hldiagr}
\begin{eqnarray}
 H_{diag}^{(i,1)(i+1,0)} \{\sigma_{r \ne i,i+1}^z\}&&  =\frac{1}{2^2} Tr_{\sigma_i,\sigma_i+1} (\sigma_{i}^z H_{diag} \{\sigma_r^z\}  )\nonumber \\ 
  = \sum_{a_1=0,z} ... && \sum_{a_{i-1}=0,z}\sum_{a_{i+2}=0,z}...  \sum_{a_N=0,z}
H_{a_1,..,a_{i-1},a_i=1,a_{i+1}=0,a_{i+2}...a_N } (l) \sigma_1^{(a_1)}    ...\sigma_{i-1}^{(a_{i-1})} \sigma_{i+2}^{(a_{i+2})}...  \sigma_N^{(a_N)} 
\nonumber \\ 
 H_{diag}^{(i,0)(i+1,1)}  \{\sigma_{r \ne i,i+1}^z\} &&  =\frac{1}{2^2} Tr_{\sigma_i,\sigma_i+1} (\sigma_{i+1}^z H_{diag} \{\sigma_r^z\}  ) \nonumber \\ 
  = \sum_{a_1=0,z} ...&& \sum_{a_{i-1}=0,z}\sum_{a_{i+2}=0,z}...  \sum_{a_N=0,z}
H_{a_1,..,a_{i-1},a_i=0,a_{i+1}=1,a_{i+2}...a_N } (l) \sigma_1^{(a_1)}    ...\sigma_{i-1}^{(a_{i-1})} \sigma_{i+2}^{(a_{i+2})}...  \sigma_N^{(a_N)} 
\label{dv2spinzauxrappel}
\end{eqnarray}

\subsection { Conservation of the invariant $I_2$ by the flow }

Since we have made the approximation that one could neglect the terms of Eq. \ref{flowtodanonclosed}, it is important to consider the dynamics of the exact invariant $I_2$
via the approximated closed flow of Eqs \ref{flowtodahdiag} and \ref{flowtodajiz}.

For the Hamiltonian of Eq. \ref{htoda}, one obtains 
\begin{eqnarray}
I_2(l) && \equiv \frac{1}{2^N}Tr(H^2(l)) = \frac{1}{2^N}
Tr \left(  H_{diag}\{\sigma_r^z\} + \sum_n
{\cal J}_{n}\{\sigma_{r \ne (n,n+1)}^z\}  ( \sigma_n^+ \sigma_{n+1}^- +\sigma_n^- \sigma_{n+1}^+ ) \right)^2
\nonumber \\
&&=\frac{1}{2^N} Tr \left( H_{diag}^2 (\sigma_r^z) + \sum_n
  \frac{  {\cal J}_{n}^2(\sigma_r^z) }{2}   \right)
\label{i2z}
\end{eqnarray}

Using Eq. \ref{flowtodahdiag},
 the diagonal contribution evolves according to
\begin{eqnarray}
I_2^{diag}(l) && =\frac{1}{2^N} Tr \left( H_{diag}^2 (\sigma_r^z)  \right)
\label{i2diagz}
\end{eqnarray}
evolves according to
\begin{eqnarray}
\frac{d I_2^{diag}(l)}{dl} && 
=\frac{2}{2^N}  Tr \left(  \frac{d H_{diag} (\sigma_r^z)}{dl} H_{diag} (\sigma_r^z)\right)
= \frac{2}{2^N} \sum_i  Tr \left(    {\cal J}_{i}^2\{\sigma_{r \ne (i,i+1)}^z\}
(\sigma_i^z - \sigma_{i+1}^z ) H_{diag}  (\sigma_r^z)  \right)
\label{i2diagzderi}
\end{eqnarray}
Since the ${\cal J}_{i}^2\{\sigma_{r \ne (i,i+1)}^z\} $ does not depend on the two spins
$\sigma_i$ and $\sigma_{i+1}$, the partial trace over these two spins alone can be
evaluated with the expansion of $H_{diag}  (\sigma_r^z) $ of Eq. \ref{dv2spinshdiag}
yielding
\begin{eqnarray}
 Tr_{\sigma_i,\sigma_{i+1}} \left( (\sigma_i^z - \sigma_{i+1}^z ) H_{diag}  (\sigma_r^z)  \right)
&&=  Tr_{\sigma_i,\sigma_{i+1}} \left( (\sigma_i^z H_{diag}  (\sigma_r^z)- \sigma_{i+1}^z H_{diag}  (\sigma_r^z) \right)
\nonumber \\ 
&& = 4  H_{diag}^{(i,1)(i+1,0)} \{\sigma_{r \ne i,i+1}^z\}  - 4 H_{diag}^{(i,0)(i+1,1)} \{\sigma_{r \ne i,i+1}^z\} 
\nonumber \\ 
&& = Tr_{\sigma_i,\sigma_{i+1}} \left(H_{diag}^{(i,1)(i+1,0)} \{\sigma_{r \ne i,i+1}^z\}  -  H_{diag}^{(i,0)(i+1,1)} \{\sigma_{r \ne i,i+1}^z\}  \right)
\label{partialtrace}
\end{eqnarray}
so that Eq. \ref{i2diagzderi} can be rewritten as
\begin{eqnarray}
\frac{d I_2^{diag}(l)}{dl} && 
= \frac{2}{2^N} \sum_i  Tr \left(    {\cal J}_{i}^2\{\sigma_{r \ne (i,i+1)}^z\}
\left(H_{diag}^{(i,1)(i+1,0)} \{\sigma_{r \ne i,i+1}^z\}  -  H_{diag}^{(i,0)(i+1,1)} \{\sigma_{r \ne i,i+1}^z\}  \right)  \right)
\label{i2diagzderifin}
\end{eqnarray}

On the other hand, using Eq. \ref{flowtodajiz},
the off-diagonal contribution in Eq. \ref{i2z}
\begin{eqnarray}
I_2^{off}(l) && =\frac{1}{2^N}\sum_n Tr \left(
  \frac{ {\cal J}_{i}^2(\sigma_r^z) }{2}   \right)
\label{i2offz}
\end{eqnarray}
evolves according to
\begin{eqnarray}
\frac{d I_2^{off}(l)}{dl} && 
= \frac{1}{2^N}\sum_n Tr \left(
   {\cal J}_{i}(\sigma_r^z) \frac{d {\cal J}_{i} \{\sigma_{k \ne (i,i+1)} ^z\}}{dl}  \right)
\nonumber \\ &&
= \frac{2}{2^N}\sum_n Tr \left(
   {\cal J}_{i}^2(\sigma_r^z) 
\left(  H_{diag}^{(i,0)(i+1,1)} \{\sigma_{r \ne i,i+1}^z\}
- H_{diag}^{(i,1)(i+1,0)} \{\sigma_{r \ne i,i+1}^z\}
  \right) \right)
\label{i2offzderi}
\end{eqnarray}
so that the sum with Eq. \ref{i2diagzderifin} yields the conservation of $I_2(l)$
of Eq. \ref{i2z}
\begin{eqnarray}
\frac{d I_2(l)}{dl} = \frac{d I_2^{diag}(l)}{dl}+\frac{d I_2^{off}(l)}{dl} =0
\label{i2conserv}
\end{eqnarray}

\subsection{ Exactness of the flow for the random XX chain with random fields }

When the initial model is the random XX chain with random fields $J^{zz}_n=0$,
one obtains that the Ansatz of Eq. \ref{htoda} is an exact solution of the flow
\begin{eqnarray}
H^{XX}(l) && = H_{diag}^{XX}\{\sigma_r^z\} + \sum_n
{\cal J}_{n}^{XX}\{\sigma_{r \ne (n,n+1)}^z\}  ( \sigma_n^+ \sigma_{n+1}^- +\sigma_n^- \sigma_{n+1}^+ )
\label{htodaxx}
\end{eqnarray}
 where the diagonal part reduces to running random fields $h_i(l)$
\begin{eqnarray}
H_{diag}^{XX}(\{\sigma_r^z\}) && = \sum_{i=1}^N  h_i(l) \sigma_i^z 
\label{XXhdiag}
\end{eqnarray}
while ${\cal J}_{n}^{XX}\{\sigma_{r \ne (n,n+1)}^z\} $ reduce to running couplings $J_n(l)$
\begin{eqnarray}
{\cal J}_{n}^{XX}\{\sigma_{r \ne (n,n+1)}^z\} = J_n(l)
\label{XXhjj}
\end{eqnarray}
so that the generation of terms with four operators in Eq. \ref{flowtodanonclosed}
identically vanishes
\begin{eqnarray}
&& \left( \frac{dH}{dl} \right)_{non-closed}  
= 0
\label{flowtodanonclosedxx}
\end{eqnarray}

The coupled flow Equations \ref{flowtodahdiag} and \ref{flowtodajiz}
reduce to
\begin{eqnarray}
\frac{d  h_i(l)}{dl}&&  = J_i^2 -  J_{i-1}^2  
\label{flowhixx}
\end{eqnarray}
and
\begin{eqnarray}
 \frac{d J_i(l) }{dl}
&&  = 2 J_i(l) \left(  h_{i+1}(l) - h_i(l) \right)
\label{flowjixx}
\end{eqnarray}
so that they coincide with the Toda flow for tridiagonal matrices 
of Eq. \ref{todatridiag} via the identification $H_{i,i+1}=J_i$ and $H_{i,i}=2h_i$.
This equivalence is consistent with the tridiagonal matrix in the fermion language
 of Eq. \ref{XXZfermions} that has to be diagonalized to obtain the free-fermions eigenvalues.
So here the Toda flow leads directly to the diagonal final result of Eq. \ref{XXflotinfinity} within the spin language,
without going through the explicit Jordan-Wigner transformation onto free-fermions operators
(Eq \ref{jordanwigner}).

\subsection{ Flow Equations for the real coefficients }

Using the expansion of Eq. \ref{jtoda}, 
 one obtains the expansion of the square
\begin{eqnarray}
{\cal J}_{n}^2\{\sigma_{r \ne (n,n+1)}^z\} && =   \sum_{a_1=0,z}.. .\sum_{a_{n-1}=0,z}\sum_{a_{n+2}=0,z} ...  \sum_{a_N=0,z}
J_n(a_1,..,a_{n-1},a_{n+2}..,a_N;l) 
\nonumber \\
&&   \sum_{b_1=0,z}.. .\sum_{b_{n-1}=0,z}\sum_{b_{n+2}=0,z} ...  \sum_{b_N=0,z}
J_n(b_1,..,b_{n-1},b_{n+2}..,b_N;l) 
\nonumber \\ 
&&  ( \sigma_1^{(a_1)}\sigma_1^{(b_1)}) ... (\sigma_{n-1}^{(a_{n-1})} \sigma_{n-1}^{(b_{n-1})} )
( \sigma_{n+2}^{(a_{n+2})}  \sigma_{n+2}^{(b_{n+2})} ) ...  (\sigma_N^{(a_N)} \sigma_N^{(b_N)} )
\label{jtodasquare}
\end{eqnarray}
so that the flow Equation \ref{flowtodahdiag} becomes
\begin{eqnarray}
\frac{dH_{diag}\{ \sigma^z_k \}}{dl}
&&  = \sum_{n}  {\cal J}_{n}^2\{\sigma_{r \ne (n,n+1)}^z\}
(\sigma_n^z - \sigma_{n+1}^z )
\nonumber \\ 
&&  = \sum_{n}  \sum_{a_1=0,z}.. .\sum_{a_{n-1}=0,z}\sum_{a_{n+2}=0,z} ...  \sum_{a_N=0,z}
J_n(a_1,..,a_{n-1},a_{n+2}..,a_N;l) 
\nonumber \\
&&   \sum_{b_1=0,z}.. .\sum_{b_{n-1}=0,z}\sum_{b_{n+2}=0,z} ...  \sum_{b_N=0,z}
J_n(b_1,..,b_{n-1},b_{n+2}..,b_N;l) 
\nonumber \\ 
&&  ( \sigma_1^{(a_1)}\sigma_1^{(b_1)}) ... (\sigma_{n-1}^{(a_{n-1})} \sigma_{n-1}^{(b_{n-1})} )
(\sigma_n^z - \sigma_{n+1}^z )
( \sigma_{n+2}^{(b_{n+2})} ) ...  (\sigma_N^{(a_N)} \sigma_N^{(b_N)} )
\label{flowtodahdiagexp}
\end{eqnarray}
Using the properties of the Pauli matrices $\sigma^{(0)}=Id$ and $\sigma_i^z$
\begin{eqnarray}
\sigma_i^{(a_i)} \sigma_i^{(b_i)} = \delta_{a_i,b_i} \sigma_i^{(0)} + (1-\delta_{a_i,b_i})\sigma_i^{z}
\label{proppauli}
\end{eqnarray}
this can be rewritten as flow equations for the $2^N$ real coefficients
of the diagonal part of Eq. \ref{hldiagr} as
\begin{eqnarray}
\frac{ dH_{c_1...c_N } (l) }{dl} && =  \sum_{n}  \sum_{a_1=0,z}.. .\sum_{a_{n-1}=0,z}\sum_{a_{n+2}=0,z} ...  \sum_{a_N=0,z}
J_n(a_1,..,a_{n-1},a_{n+2}..,a_N;l) 
\nonumber \\
&&   \sum_{b_1=0,z}.. .\sum_{b_{n-1}=0,z}\sum_{b_{n+2}=0,z} ...  \sum_{b_N=0,z}
J_n(b_1,..,b_{n-1},b_{n+2}..,b_N;l) 
\nonumber \\ 
&& (\delta_{c_n,1} \delta_{c_{n+1},0} -\delta_{c_n,0} \delta_{c_{n+1},1} )
\prod_{i \ne (n,n+1)}  \left( \delta_{c_i,0} \delta_{a_i,b_i}  + \delta_{c_i,z}(1-\delta_{a_i,b_i}) \right)
\label{hldiagflowcoef}
\end{eqnarray}

For instance the flow equations for the random fields $h_i(l)$ 
with only a single non-vanishing index $c_i=z$ reads
\begin{eqnarray}
&& \frac{ dh_i (l) }{dl} = \frac{ dH_{c_1=0,.. c_{i-1}=0,,c_i=1, c_{i+1}=0,.....c_N=0 } (l) }{dl} 
\label{hldiagflowcoefhi} \\ 
&&
 =   \sum_{a_1=0,z}.. .\sum_{a_{i-2}=0,z}\sum_{a_{i+2}=0,z} ...  \sum_{a_N=0,z}
\left(\sum_{a_{i-1}=0,z}  J_i^2(a_1,..,a_{i-1},a_{i+2}..,a_N;l)- \sum_{a_{i+1}=0,z}  J_{i-1}^2(a_1,..,a_{i-2},a_{i+1}..,a_N;l) 
    \right) 
\nonumber 
\end{eqnarray}

Similarly, the flow Eq. \ref{flowtodajiz} can be expanded into coefficients using Eqs
\ref{jtoda} and \ref{dv2spinzauxrappel}
\begin{eqnarray}
 \frac{d {\cal J}_{n} \{\sigma_{k \ne (n,n+1)} ^z\}}{dl}
&&  = 2 {\cal J}_{n}\{\sigma_{k \ne (n,n+1)} ^z\}
\left(  H_{diag}^{(n,0)(n+1,1)} \{\sigma_{r \ne n,n+1}^z\}
- H_{diag}^{(n,1)(n+1,0)} \{\sigma_{r \ne n,n+1}^z\}
\right)
\nonumber \\ 
&&  = 2 
\sum_{a_1=0,z} ...  \sum_{a_{i-1}=0,z}\sum_{a_{i+2}=0,z}...  \sum_{a_N=0,z}H_{a_1,..,..a_N } (l)
(\delta_{a_n,0} \delta_{a_{n+1},1} -\delta_{a_n,1} \delta_{a_{n+1},0} )
\nonumber \\ 
&& \sum_{b_1=0,z}.. .\sum_{b_{n-1}=0,z}\sum_{b_{n+2}=0,z} ...  \sum_{b_N=0,z}
J_n(b_1,..,b_{n-1},b_{n+2}..,b_N;l) 
 \nonumber \\  &&
  ( \sigma_1^{(a_1)}\sigma_1^{(b_1)}) ... (\sigma_{n-1}^{(a_{n-1})} \sigma_{n-1}^{(b_{n-1})} )
(  \sigma_{n+2}^{(a_{n+2})}  \sigma_{n+2}^{(b_{n+2})} ) ...  (\sigma_N^{(a_N)} \sigma_N^{(b_N)} )
\label{flowtodajizexpansion}
\end{eqnarray}

Using again Eq. \ref{proppauli}
this can be rewritten as flow equations for the real coefficients
of Eq. \ref{jtoda} as
\begin{eqnarray}
\frac{dJ_n(c_1,..,c_{n-1},c_{n+2}..,c_N;l)}{dl}&&  =2 
\sum_{a_1=0,z} ...  \sum_{a_{i-1}=0,z}\sum_{a_{i+2}=0,z}...  \sum_{a_N=0,z}H_{a_1,..,..a_N } (l)
(\delta_{a_n,0} \delta_{a_{n+1},1} -\delta_{a_n,1} \delta_{a_{n+1},0} )
\nonumber \\ 
&& \sum_{b_1=0,z}.. .\sum_{b_{n-1}=0,z}\sum_{b_{n+2}=0,z} ...  \sum_{b_N=0,z}
J_n(b_1,..,b_{n-1},b_{n+2}..,b_N;l) 
 \nonumber \\  &&
\prod_{i \ne (n,n+1)}  \left( \delta_{c_i,0} \delta_{a_i,b_i}  + \delta_{c_i,z}(1-\delta_{a_i,b_i}) \right)
\label{jtodacoef}
\end{eqnarray}

For instance the flow equations for the $J_n(l)$ 
where all indices vanish $c_i=0$ reads
\begin{eqnarray}
\frac{dJ_n(l) }{dl}&& = \frac{dJ_n(c_1=0,..,c_{n-1}=0,c_{n+2}=0..,c_N=0;l)}{dl}
 \label{jtodacoefex} \\ &&   =2 
\sum_{a_1=0,z} ...  \sum_{a_{i-1}=0,z}\sum_{a_{i+2}=0,z}...  \sum_{a_N=0,z}H_{a_1,..,..a_N } (l)
(\delta_{a_n,0} \delta_{a_{n+1},1} -\delta_{a_n,1} \delta_{a_{n+1},0} )
J_n(a_1,..,a_{n-1},a_{n+2}..,a_N;l) 
\nonumber
\end{eqnarray}

\subsection{ Transformation of the spin operators }

Up to now we have focused only on the flow of the Hamiltonian, but it is of course
interesting to consider the flow of other observables $A$ via the flow equation
analogous to Eq. \ref{deriHl}
\begin{eqnarray}
\frac{dA(l)}{dl}= \left[ \eta(l), A(l) \right]
\label{flowA}
\end{eqnarray}

With the choice of the Toda generator of Eq. \ref{todachoicespin2},
the creation operators evolve according to
\begin{eqnarray}
 \frac{d \sigma_n^+}{dl}  = [ \eta(l) ,\sigma_n^+ ] 
&& =
\sum_{i} [  {\cal J}_{i} \{\sigma_{k \ne (i,i+1)} ^z\} ( \sigma_i^+ \sigma_{i+1}^- - \sigma_i^- \sigma_{i+1}^+ ) , \sigma_n^+ ] 
\nonumber \\ && 
=  \left(  {\cal J}_{n} \{\sigma_{k \ne (n,n+1)} ^z\}  \sigma_n^z \right)  
\sigma_{n+1}^+
 - \left(    {\cal J}_{n-1} \{\sigma_{k \ne (n-1,n)} ^z\} \sigma_n^z \right) \sigma_{n-1}^+
\nonumber \\ && +2 \sum_{i\ne (n-1,n)}   {\cal J}_{i}^{(n,1)} \{\sigma_{k \ne (i,i+1)} ^z\} \sigma_n^+  ( \sigma_i^+ \sigma_{i+1}^- - \sigma_i^- \sigma_{i+1}^+ )
\label{si1p} 
\end{eqnarray}
The annihilation operators evolve similarly as
\begin{eqnarray}
\frac{d \sigma_n^-}{dl} && =  \left(  {\cal J}_{n} \{\sigma_{k \ne (n,n+1)} ^z\}  \sigma_n^z \right)  
\sigma_{n+1}^-
 - \left(    {\cal J}_{n-1} \{\sigma_{k \ne (n-1,n)} ^z\} \sigma_n^z \right) \sigma_{n-1}^-
\\ \nonumber && - 2 \sum_{i\ne (n-1,n)}   {\cal J}_{i}^{(n,1)} \{\sigma_{k \ne (i,i+1)} ^z\} \sigma_n^-  ( \sigma_i^+ \sigma_{i+1}^- - \sigma_i^- \sigma_{i+1}^+ )
\label{si1m} 
\end{eqnarray}
while the operators $\sigma_n^z $ evolve as
\begin{eqnarray}
\frac{d \sigma_n^z}{dl} && = [ \eta,\sigma_n^z ] =
\sum_{i} [  {\cal J}_{i} \{\sigma_{k \ne (i,i+1)} ^z\} ( \sigma_i^+ \sigma_{i+1}^- - \sigma_i^- \sigma_{i+1}^+ ) , \sigma_n^z ] 
\nonumber \\ && 
= 2 {\cal J}_{n-1} \{\sigma_{k \ne (n-1,n-1+1)} ^z\}
  ( \sigma_{n-1}^+ \sigma_{n}^- + \sigma_{n-1}^- \sigma_{n}^+ )  
-2 {\cal J}_{n} \{\sigma_{k \ne (n,n+1)} ^z\}  ( \sigma_n^+ \sigma_{n+1}^- + \sigma_n^- \sigma_{n+1}^+ )
\label{si1z}
\end{eqnarray}

\subsection{ Discussion on the numerical implementation }

For the diagonalization of matrices, the Jacobi method based on discrete rotations is not considered as the optimal numerical method nowadays, and has been replaced by the QR-algorithm or the Toda flow that can be seen as its continuous counterpart. 
As a consequence, for the Many-Body-Localization problem, 
one could also expect that the discret scheme \cite{rademaker}
which is equivalent to the Jacobi method
 might be outperformed in the future by more efficient numerical strategies.
In this section,  we have proposed to use an analog of the Toda flow for the XXZ chain with random fields..
We believe that the flow Equations \ref{hldiagflowcoef}
and \ref{jtodacoef} for the real coefficients can be used 
to study various truncations in the number of non-vanishing indices for the
most relevant couplings, and can be studied to higher orders than in the discrete 
scheme \cite{rademaker}.

However this numerical implementation is clearly beyond the scope of the present manuscript and beyond the numerical ability of the present author. For readers
interested into this type of numerical implementation, we feel that the more relevant references could be the Physics reference \cite{white} dedicated to the numerical applications of both the discrete and the continuous schemes with their comparison, and the Applied Mathematics reference \cite{calvo} focusing of the numerical solutions of continuous isospectral flows. These previous achievements at least show that there is no insurmountable practical problem in these numerical continuous flows.
Finally Reference \cite{rademaker} contains many discussions on the practial application of the discrete scheme that are more specific to the Many-Body-Localization problem
and on the various observables that are interesting to study.

\section{ Conclusion }

\label{sec_conclusion}

In this paper, we have revisited the various flows towards diagonalization that 
have been introduced in the past, in order to stress the freedom in the choices that can be made :

(i) discrete or continuous formulation

(ii) strategy based on the steady decay of $I_2^{off}$ or on the repeated application of the Hamiltonian

(iii) schemes that tend to generate all possible couplings in the running Hamiltonian or schemes that are able to preserve some 'sparsity' of the initial condition.

We have then focused on the random XXZ chain with random fields in order to determine the best closed flow within the subspace of running Hamiltonians containing only two ladder operators on consecutive sites. For the special case of the free-fermion random XX chain with random fields, we have shown that the flow coincides with the Toda differential flow for tridiagonal matrices which can be seen as the continuous analog of the discrete QR-algorithm. For the random XXZ chain with random fields that displays a Many-Body-Localization transition, the present differential flow is an interesting alternative to the discrete flow that has been proposed recently to study the Many-Body-Localization properties in a model of interacting fermions \cite{rademaker}. We hope that the differential flow for the XXZ-chain can be transformed into an efficient numerical procedure where the effects of the order of truncation can be systematically studied up to higher orders. 
Finally, since the necessity to diagonalize matrices and operators appears almost everywhere
in sciences, we hope that the idea of the Toda flow can be adapted in various fields.


\begin{thebibliography}{99}

\bibitem{revue_huse}
R. Nandkishore and D. A. Huse, Ann. Review of Cond. Mat. Phys. 6, 15 (2015).

\bibitem{revue_altman}
 E. Altman and R. Vosk, Ann. Review of Cond. Mat. Phys. 6, 383 (2015).


\bibitem{jacobi}
C.G.J. Jacobi, Crelle's Journal (in german) 30, 51 (1846).

\bibitem{white}
S.R. White, J. Chem. Phys. 117, 7472 (2002).

\bibitem{rademaker}
L. Rademaker and M. Ortuno, Phys. Rev. Lett. 116, 010404 (2016).

\bibitem{wegner_first}
F. Wegner, Annalen der Physik 3, 77 (1994).

\bibitem{wegner_reviews}
F. Wegner, Phys. Rep. 348, 77 (2001) ; \\
F. Wegner, J. Phys. A Math. Gen. 39, 8221 (2006); \\
F. Wegner, Int. J. of Mod. Phys. A 29, 143043 (2014).

\bibitem{book_kehrein}
S. Kehrein, ``The flow equation approach to many-particle systems'', Springer-Verlag Berlin (2006). 

\bibitem{wilson}
S.D. Glazek and K.G. Wilson, Phys. Rev. D 48, 5863 (1993) ; \\
S.D. Glazek and K.G. Wilson, Phys. Rev. D 49, 4214 (1994) ; \\
S.D. Glazek, Int. J. of Mod. Phys. A 30, 1530023 (2015).

\bibitem{math1}
M.T. Chu and K. Driessel, Siam J. Num. Anal. 1050 (1990).

\bibitem{math2}
R.W. Brockett, Lin. Alg. and Appl. 146, 79 (1991).

\bibitem{book_krylov} Y. Saad,
``Iterative methods for sparse linear systems'' ,
SIAM (2003).

\bibitem{henon}
M. Henon, Phys. Rev. B 9, 1921 (1974).

\bibitem{flaschka}
H. Flaschka, Phys. Rev. B 9, 1924 (1974).

\bibitem{moser}
J. Moser, in Dynamical Systems theory and Applications, Springer Verlag, Berlin 1975.

\bibitem{deift}
P Deift, T. Nanda and C. Tomei, SIAM J. 20, 1 (1983).

\bibitem{symes}
W.W. Symes, Physica 1D, 339 (1980).

\bibitem{watkins}
D.S. Watkins, SIAM 26, 379 (1984).

\bibitem{chu}
M.T. Chu, SIAM 30, 375 (1988)

\bibitem{elsner}
D.S. Watkins, SIAM J. Matrix Anal. Appl. 11, 301 (1990).

\bibitem{tomei}
C. Tomei, arxiv:1508.03229.

\bibitem{toda}
M. Toda, ``Theory of nonlinear lattices'', Springer Verlag Berlin (1989).

\bibitem{mielke_band}
A. Mielke, Eur. Phys. J. B 5, 605 (1998).



\bibitem{atas}
Y.Y. Atas and E. Bogomolny, J. Phys. A Math. Theor. 47, 335201 (2014) ; \\
Y.Y. Atas and E. Bogomolny, arxiv:1503.04508.

\bibitem{keating}
J.P. Keating, N. Linden and H.J. Wells, arxiv:1403.1114 ; 
J.P. Keating, N. Linden and H.J. Wells, Comm. Math. Phys. 338, 81 (2015).


\bibitem{emergent_swingle}
B. Swingle, arxiv:1307.0507.

\bibitem{emergent_serbyn}
M. Serbyn, Z. Papic and D.A. Abanin, Phys. Rev. Lett. 111, 127201 (2013).

\bibitem{emergent_huse}
D.A. Huse, R. Nandkishore and V. Oganesyan, Phys. Rev. B 90, 174202 (2014).

\bibitem{emergent_ent}
A. Nanduri, H. Kim and D.A. Huse, Phys. Rev. B 90, 064201 (2014).

\bibitem{imbrie}
J. Z. Imbrie,  J. Stat. Phys. 163, 998 (2016).

\bibitem{serbyn_quench}
M. Serbyn, Z. Papic and D.A. Abanin, Phys. Rev. B 90, 174302 (2014).

\bibitem{emergent_vidal}
A. Chandran, I.H. Kim, G. Vidal and  D.A. Abanin, Phys. Rev. B 91, 085425 (2015).

\bibitem{emergent_ros}
V. Ros, M. M\"uller and A. Scardicchio, Nucl. Phys. B 891, 420 (2015).


\bibitem{rsrgx}
D. Pekker, G. Refael, E. Altman, E. Demler and V. Oganesyan,
Phys. Rev. X 4, 011052 (2014).

 \bibitem{rsrgx_moore}
Y. Huang and J.E. Moore, Phys. Rev. B 90, 220202(R) (2014).

\bibitem{vasseur_rsrgx}
R. Vasseur, A. C. Potter and S.A. Parameswaran, Phys. Rev. Lett. 114, 217201 (2015).


\bibitem{yang_rsrgx}
M. Pouranvari and K. Yang, Phys. Rev. B 92, 245134 (2015).

\bibitem{rsrgx_bifurcation}
 Y.Z. You, X.L. Qi and C. Xu, Phys. Rev. B 93, 104205 (2016).




\bibitem{emergent_c}
C. Monthus,  J. Stat. Mech. 033101 (2016).



\bibitem{ma_dasgupta}
S.K. Ma, C. Dasgupta and C.K. Hu, Phys. Rev. Lett. 43, 1434 (1979);\\
C. Dasgupta and S.K. Ma, Phys. Rev. B 22, 1305 (1980).

\bibitem{fisher_AF}
D. S. Fisher, Phys. Rev. B 50, 3799 (1994).

\bibitem{fisher}
D. S. Fisher, Phys. Rev. Lett. 69, 534 (1992); \\
D. S. Fisher, Phys. Rev. B 51, 6411 (1995).


\bibitem{fisherreview}
D. S. Fisher, Physica A 263, 222 (1999).

\bibitem{strong_review}
F. Igloi and C. Monthus, Phys. Rep. 412, 277 (2005).

\bibitem{refael_review}
G. Refael and E. Altman, Comptes Rendus Physique, Vol. 14, 725 (2013).


\bibitem{alet}
D. J. Luitz, N. Laflorencie and F. Alet, Phys. Rev. B 91, 081103 (2015).



\bibitem{alet_dyn}
D. J. Luitz, N. Laflorencie and F. Alet, 
Phys. Rev. B 93, 060201 (2016).


\bibitem{luitz}
D. Luitz, Phys. Rev. B 93, 134201 (2016).




\bibitem{Wolff-Sreview}
S. Braavyi, D.P. DiVincenzo and D. Loss, Ann. Phys. 326, 2793 (2011)

\bibitem{refael_fisher}
G. Refael and D.S. Fisher, Phys. Rev. B 70, 064409 (2004).


\bibitem{WolffS_LR}
P. Hauke and M. Heyl, Phys. Rev. B 92, 134204 (2015).



\bibitem{choixsigne}
T.D. Morris, N.M. Parzuchowski and S.K. Bogner, Phys. Rev. C 92, 034331 (2015).

\bibitem{calvo}
M.P. Calvo, A. Iserles and A. Zanna, Math. of Comput. 66, 1461 (1997).




 \end{thebibliography}
\end{document}